\documentclass[fleqn,usenatbib,useAMS]{mnras}
\usepackage[T1]{fontenc}
\usepackage{ae,aecompl}
\usepackage{graphicx}
\usepackage{mathptmx}
\usepackage{amsmath}
\usepackage{amssymb}
\usepackage{multicol}
\usepackage{color}
\usepackage{pdfcolmk}
\usepackage{soul}
\usepackage[normalem]{ulem}

\definecolor{grn}{RGB}{0,100,0}

\newcommand\redout[1]{\bgroup\markoverwith
{\color{red}{\rule[.5ex]{2pt}{0.4pt}}}\ULon{#1}\vspace{-0.5ex}}
\newcommand\redoutt[1]{\bgroup\markoverwith
{\color{red}{\rule[1.25ex]{2pt}{0.4pt}}}\ULon{#1}\vspace{-1.25ex}}

\setstcolor{red}

\gdef\Min{${}^{\prime}$\llap{.}}

\def\etal{\hbox{et al.}}

\gdef\ltsima{$\scriptscriptstyle \; \buildrel < \over \sim \;$}
\gdef\simlt{\lower.3ex\hbox{\ltsima}}

\gdef\gtsima{$\scriptscriptstyle \; \buildrel > \over \sim \;$}
\gdef\simgt{\lower.3ex\hbox{\gtsima}}

\gdef\about{\raise.3ex\hbox{$\scriptscriptstyle \sim $}}

\title[Gravity and the Nonlinear Growth of Structure]{
Gravity and the Nonlinear Growth of Structure in the Carnegie-Spitzer-IMACS Redshift Survey
}

\author[Kelson \etal]{Daniel D.~Kelson$^{1}$, Louis E. Abramson$^1$, Andrew J. Benson$^1$, Shannon G. Patel$^1$,\newauthor
Stephen A. Shectman$^1$, Alan Dressler$^1$, Patrick J. McCarthy$^1$, John S. Mulchaey$^1$, \&\newauthor
Rik J. Williams$^2$
\\
\\
$^{1}$The Observatories, The Carnegie Institution for Science,
813 Santa Barbara St., Pasadena, CA 91101\\
$^{2}$Uber Technologies, Inc., 1455 Market St. 4th Floor, San Francisco CA 94103
}

\begin{document}
\label{firstpage}
\pagerange{\pageref{firstpage}--\pageref{lastpage}}
\maketitle

\begin{abstract}
A key obstacle to developing a satisfying theory of galaxy evolution is the difficulty in extending 
analytic descriptions of early structure formation into full nonlinearity, the regime in which galaxy 
growth occurs. Extant techniques, though powerful, are based on approximate numerical methods whose 
Monte Carlo-like nature hinders intuition building. Here, we develop a new solution to this problem 
and its empirical validation. We first derive closed-form analytic expectations for the evolution of fixed 
percentiles in the real-space cosmic density distribution, {\it averaged over representative
volumes observers can track cross-sectionally\/}. Using the Lagrangian forms of the fluid equations,
we show that percentiles in $\delta$---the density relative to the median---should grow as 
$\delta(t)\propto\delta_{0}^{\alpha}\,t^{\beta}$, where $\alpha\equiv2$ and $\beta\equiv2$ 
for Newtonian gravity at epochs after the overdensities transitioned to nonlinear growth. We then use 
9.5 sq.~deg.~of Carnegie-Spitzer-IMACS Redshift Survey data to map {\it galaxy\/} environmental densities
over $0.2<z<1.5$ ($\sim$7~Gyr) and infer $\alpha=1.98\pm0.04$ and $\beta=2.01\pm0.11$---consistent
with our analytic prediction. These findings---enabled by swapping the Eulerian domain of most work on
density growth for a Lagrangian approach to real-space volumetric averages---provide some of the strongest evidence 
that a lognormal distribution of early density fluctuations indeed decoupled from cosmic expansion to grow through 
gravitational accretion. They also comprise the first exact, analytic description of the nonlinear growth of structure
extensible to (arbitrarily) low redshift. We hope these results open the door to new modeling of,
and insight-building into, galaxy growth and its diversity in cosmological contexts.
\end{abstract}



\section{Introduction}

The growth of structure out of the
random density fluctuations at early times is a key tenet of modern cosmology \citep[e.g.][]{peebles1967}, and
the core of any theory of galaxy formation \citep[e.g.][]{blumenthal1984}. In Press-Schechter
\citep{press1974} and Extended Press-Schechter \citep{bond1991} formalisms, the growth of Fourier modes with time
follows linear theory prior
to decoupling from the Hubble expansion, with the addition of semi-analytical approximations for baryonic
physics \citep[e.g.][]{kauffmann1993,cole1994} providing nearly all modern inferences into
galaxy formation and evolution \citep[e.g.][]{benson2003}.

After decoupling from the Hubble expansion, however, density peaks transition from linear to non-linear growth.
This nonlinear evolution of the cosmological matter density field is a difficult theoretical problem as well as an
intractable observational one, owing to the impossibility of studying the time---or longitudinal---evolution of individual density
fluctuations\footnote{Within different redshift slices---i.e. at different epochs---one's observations cover independent
{\it cross sections\/} of different density fluctuations, and do not track the {\it longitudinal\/} evolution of the same
density fluctuations \citep[e.g.][]{abramson2016}.}. The long-term behaviour of the matter density field in this regime has largely
been trusted to $N$-body simulations, as accurate analytical methods for following the long-term trajectories of
individual modes or halos have not yet been discovered. Furthermore, deriving an empirical picture of the growth of
nonlinear structure requires the inversion of cross-sectional studies of distributions of different galaxies at
different epochs---a fraught and perhaps underdetermined mathematical exercise. And while it has long been recognized
that the growth of structure in the nonlinear regime is a difficult diffusion problem \citep[e.g.,][]{bond1991}, the
non-Markovian\footnote{When random changes to a process---such as the motions of particles in an ideal gas---are uncorrelated with previous changes, the process is said to be {\it Markovian}, of which Brownian motion is an example. When changes to a process are correlated over time, it is said to be {\it non-Markovian}.} nature of halo growth trajectories complicates the derivation of mathematical expectations for the
behaviour of mass accretion beyond those from the central limit theorem or perturbation theory
\cite[e.g.,][]{coles1991,ma1995,ma2000,achitouv2013}.

As it turns out, simple solutions to the equations governing the mean growth of density fluctuations are possible if they're
recast in terms of {\it ensembles\/} or {\it percentiles\/} in a representative distribution.
Doing so conveniently shifts the fluid equations from the
standard---and complicated---Eulerian framework to a simpler, Lagrangian one, allowing us to derive new analytical
expectations for the mean growth of densities in the nonlinear regime.
Lagrangian {\it perturbative\/} approaches to the growth of structure
have been used for decades \citep[e.g.][]{zeldovich1970,moutarde1991,bertschinger1992,bertschinger1994,bouchet1996}
to provide insights into the topology and geometry of those first structures to collapse, as well as
more accurate approximations for the rates at which structure grows while they remain attached to the Hubble
expansion. In contrast to linear theory, or Lagrangian perturbation theory, our work does not begin
with the assumption of $\delta \ll 1$ while linearizing the equations of motion or expanding them to higher order powers
of $\delta$. Instead, we look at the mean behavior of fluctuations that have overcome the Hubble
expansion by volume-averaging the equations of motion, with a specific eye towards understanding
the dependence of mean growth trajectories on their initial conditions.

To validate the new analytical predictions---and the underlying picture of nonlinear growth
arising from the gravitational collapse of modes that have decoupled from the Hubble
expansion---we use the Carnegie-Spitzer-IMACS (CSI) Redshift Survey, a data set uniquely suited
to testing this framework. CSI's combination of broadband photometry and low-dispersion
spectroscopy enabled the
measurement and inference of a range of derived properties, such as redshifts, stellar masses, emission line
luminosities, and information on recent star-formation. As discussed in \citet{kelson2014b} the survey was designed to
study the evolution of galaxies, their environments, and the interplay between them. By selecting in the near-IR, the
CSI Survey efficiently traces the stellar mass of average galaxies to $z\sim 1.5$ (and more massive ones up to $z\sim 1.8$).

\begin{figure*}
\centerline{
\includegraphics[width=0.40\textwidth]{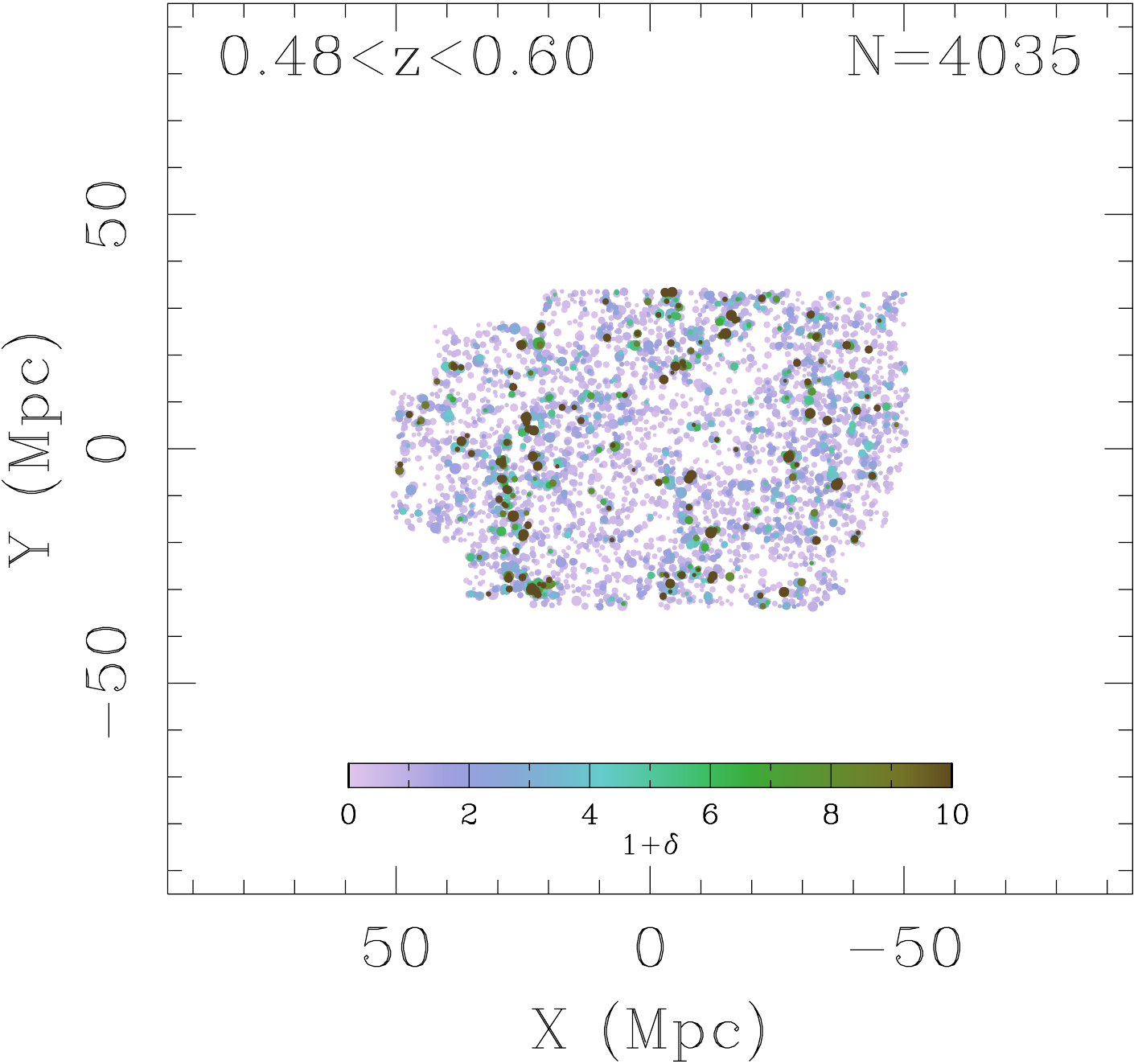}
\includegraphics[width=0.40\textwidth]{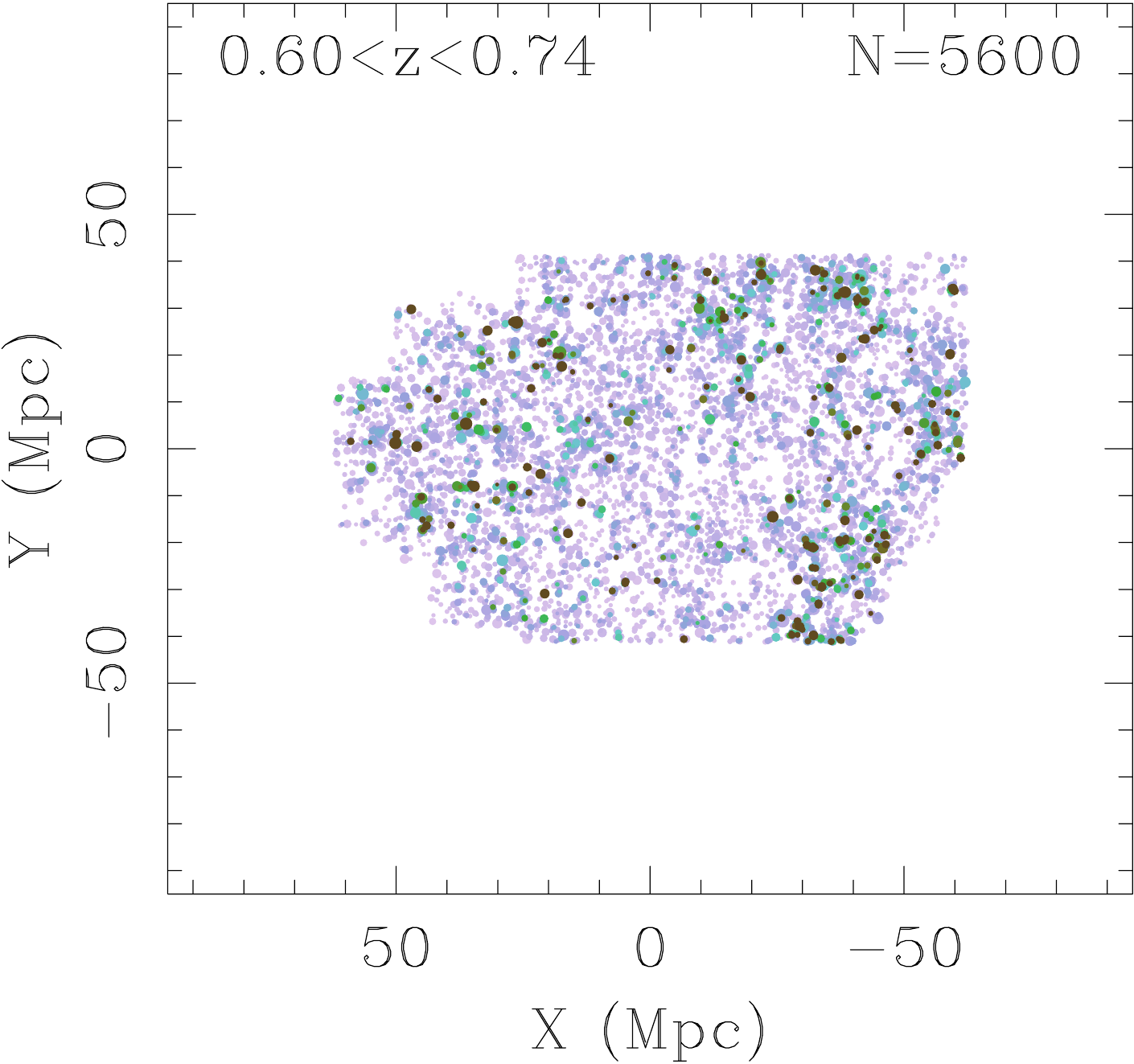}}
\bigskip
\centerline{
\includegraphics[width=0.40\textwidth]{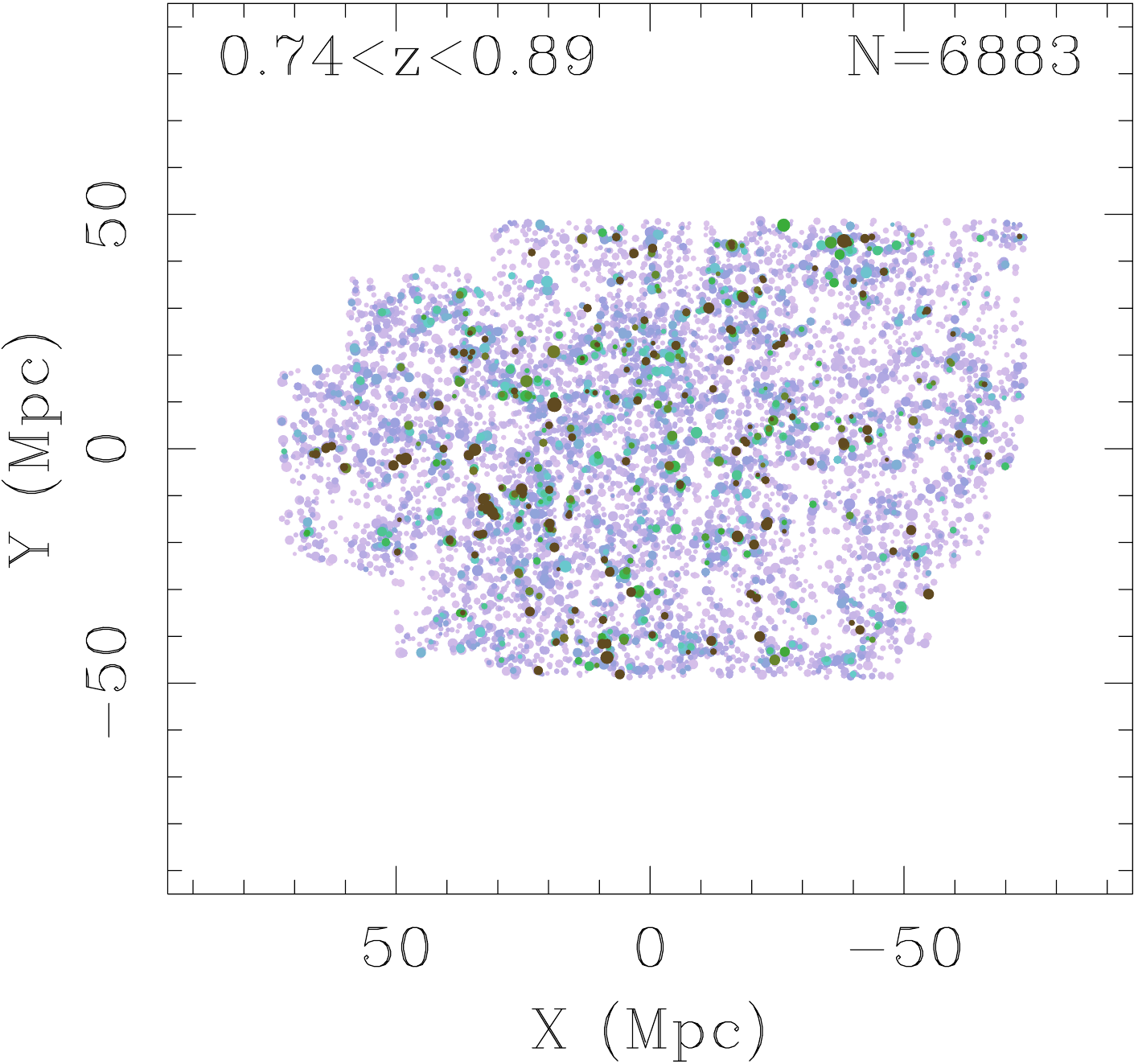}
\includegraphics[width=0.40\textwidth]{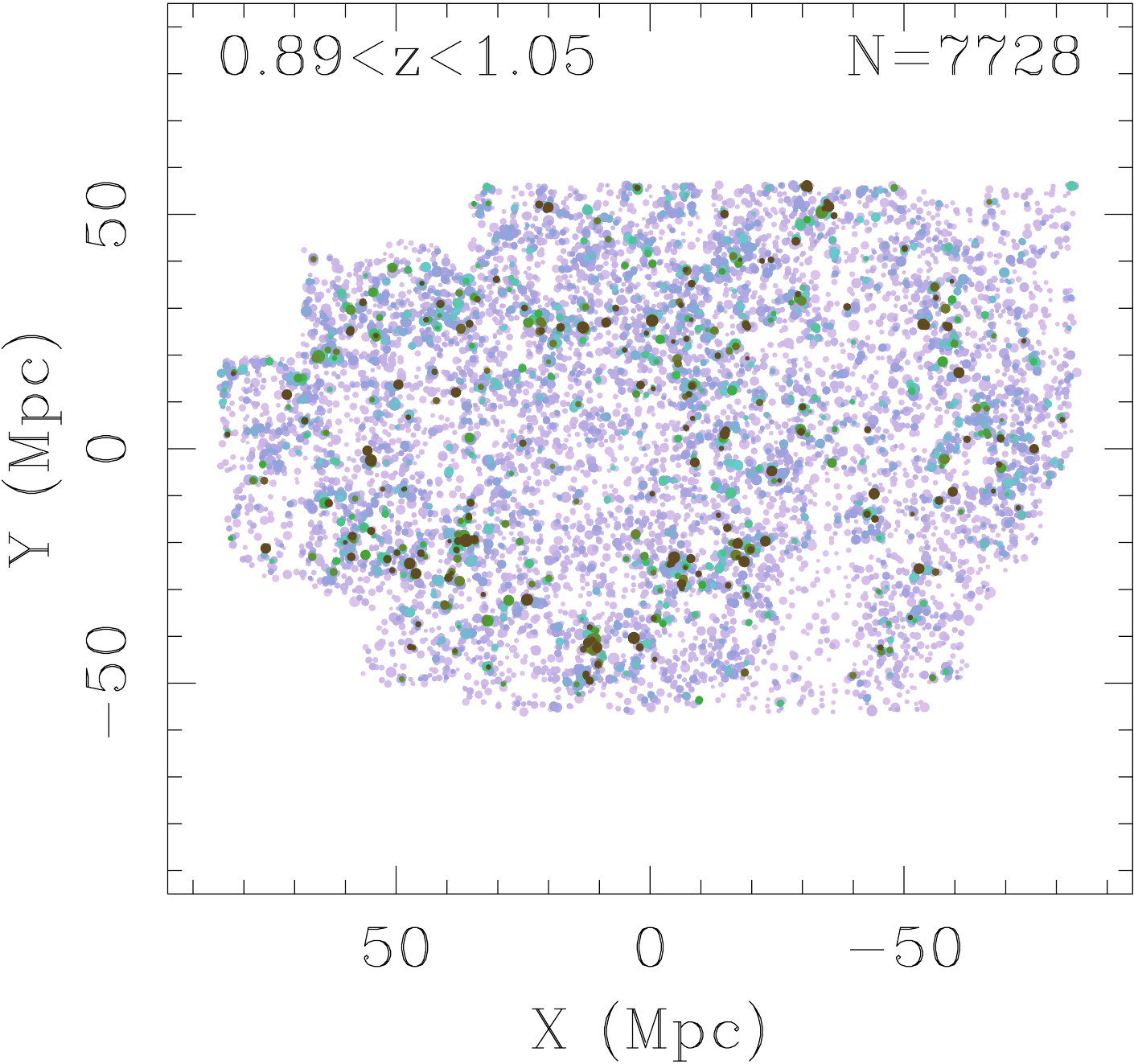}}
\caption{The positions, in comoving Mpc. for galaxies with stellar masses $M_*\ge 10^{10} M_\odot$
in four redshift slices in the CSI XMM field. The colour of each point reflects the local stellar mass density relative
to the median density in the slice (see \S \ref{sec:local}), while the point sizes reflect the stellar masses of the
individual galaxies. High density regions are expected to grow in density contrast with time, while low density regions
are expected to be hollowed out.
\label{fig:XMMslices}}
\end{figure*}

In this work we measure distributions of local stellar mass densities using CSI's $\sim 9.5$ square degrees of coverage
in the SWIRE XMM and CDFS fields. To illustrate the scale of these fields we show four redshift slices of the XMM field
in Figure \ref{fig:XMMslices}, colour coding the galaxy points by local stellar mass density
(as described in \S \ref{sec:local}). The CDF field, slightly smaller in area, shows visible structures that are similar
in characteristic to the XMM field.

Percentiles in the distribution of local stellar mass densities---over approximately 7 Gyr of cosmic time---are used as
proxies that trace the average growth of ensembles of matter density peaks. In this work we assume that local matter density
fluctuations are directly traced by stellar mass (i.e. near-IR, luminous galaxies)---at least over those epochs at which
galaxies exist, and thus use the evolution of density at fixed percentile to characterise the nonlinear growth of
structure.

By deriving analytical expectations for how density distributions should evolve {\it in the mean\/} using a Lagrangian
formalism, our comparisons with the CSI measurements then provide explicit tests that the growth of structure
is indeed a process of gravitational collapse. Should the accuracy of the derived algebraic forms be borne out, studies
of the growth of structure and the growth of galaxies may then be better understood---or even more accurately
modeled---through similar analytical approaches.

The paper is structured as follows: first, we derive in \S \ref{sec:frame} how ensembles of early overdensities that
decoupled from the Hubble expansion should evolve on average.
In \S \ref{sec:data}, the CSI data set is summarized, including upgrades to the SED modeling and
resulting improvements to survey robustness and data quality. In \S \ref{sec:local} we describe measuring local
densities using Delaunay triangulation, followed by an empirical picture of the evolution in densities over cosmic time
in \S \ref{sec:evolution}. From the observed evolution we infer the form of the initial distrbution of densities at the
time when stellar mass growth began in the universe in \S \ref{sec:info}, to be used in \S \ref{sec:global} to fit for
the mean growth of  overdensities (and diminution of underdensities) as power-laws in both time and initial density simultaneously. Implications of
our work---which confirm the analytical forms derived in \S \ref{sec:frame}---are summarized in \S \ref{sec:imps}.

The cosmological parameters used in this work are $h=0.68$, $\Omega_M=0.31$, and $\Omega_\Lambda=0.69$ \citep{planck2015}.


\section{Expectations for the Growth of Structure in the Real Domain}
\label{sec:frame}

The continuity equation dictates the evolution of individual density fluctuations with time under the constraint of mass
conservation. However, the integration of the fluid equations---a procedure without closed form solutions---is only
truly of astronomical utility when volume-averaged. From an {\it astroinformatic\/} perspective, observations cannot
provide longitudinal studies of individual parcels of mass, because one observes cross sections of different density
fluctuations at different epochs.

Operationally---and from the perspective of observers---this apparent shortcoming turns out to be a strength, as
expectation values can be derived straightforwardly for well-defined cross-sectional studies of density fluctuations.
The ``trick'' here is in recognizing that for regions that have decoupled from the Hubble expansion,
Gauss's theorem simplifies the continuity equation in very helpful ways when volume-averaging both sides. We note that
the procedure of averaging the fluid equations---temporally or spatially---is occasionally referred to as a ``Reynold's
decomposition.''

We now derive the mean Lagrangian time derivative of $\delta$, where $\delta \equiv \rho/\langle\rho\rangle -1$,
for fixed percentiles in $\delta$, and where $\rho$ is the matter density at arbitrary locations within regions
that have transitioned to nonlinear growth.
In tracking the evolution of the distribution of local stellar mass densities, we
make the assumption that stellar mass growth arises through the accretion of baryons--either in the form of gas to be
converted to stars, or in the form of stars already converted from gas through previous accretion events---with the
stellar baryons in either case tracking dark matter accretion {\it in the mean\/}. This assumption
is only being made on scales much larger than that of individual galaxy-scale halos, through the summing of stellar
mass over the many neighbouring vertices in the Delaunay triangulation (see below).
Thus any galaxy-scale halo-mass-dependence of the efficiency with which baryons are converted to stars is averaged over
many galaxies at a time.

We start with the Lagrangian continuity equation already cast in terms of overdensity $\delta$: In this approach
$\delta$---normally a function of space and time---is now a function of an abstract mass-parcel coordinate, ${\mathbf m}$, and
conformal time, $\tau\equiv t/a$:
\begin{equation}
\frac{D\delta({\mathbf m},\tau)}{D\tau} = -[1+\delta({\mathbf m},\tau)] [\nabla \cdot {\mathbf u}({\mathbf m},\tau)]
\label{eq:den}
\end{equation}
and, because we are going to discuss density fluctuations that have just (barely) decoupled from the Hubble expansion,
the relevant form of the momentum equation is
\begin{equation}
\frac{D{\mathbf u}({\mathbf m},\tau)}{D\tau} = -\nabla\phi
\label{eq:vel}
\end{equation}
where the gravitational potential is given by
\begin{equation}
\nabla^2\phi = 4\pi Ga^2\rho_M \delta({\mathbf m},\tau)
\label{eq:phi}
\end{equation}
and
$\rho_M\equiv (3\Omega_M H_0^2)/(8\pi G a^3)$, the mean matter density at the epoch corresponding to scale factor $a$.

The above equations can be contrasted to those used in linear theory, and perturbation theory in general, in
which one starts with the fluid
equations, simplifies them assuming $\delta\ll 1$, takes the time derivative of both sides of the continuity equation,
substitutes the momentum equation, and then analytically solves for the growth of modes still embedded within the
Hubble expansion.

The key difference here is that while traditional approaches keep the Hubble expansion term(s), the
disconnection of density fluctuations from the Hubble expansion allows one to both simplify the momentum equation as above, and, as will
be shown below, somewhat trivially apply the divergence theorem. These differences between our approach and previous work
obviate easy comparisons or connections. Once fluctuations have overcome the
Hubble expansion, they simply do not speak to the additional ``complicating'' terms of the fluid equations
that involve ${\dot a}$---and it is those terms that drive the forms of growth in, e.g., linear theory.

Even if one adopted a highly simplified ``two-phase'' schematic---with bubbles of matter collapsing after their
disconnection from the Hubble expansion while remaining surrounded by a medium continuing that expansion and behaving
according to linear theory---volume-averaged growth functions for the latter phase is not meaningful because modes grow
independently and in-place. For the moment an equivalent analytical treatment of the intermediate density domain must
remain an unsolved piece of the picture, as doing so remains beyond the scope of this initial work.

Using Equations \ref{eq:den}--\ref{eq:phi}, we now proceed to derive the volumetric mean change in density as functions of
conformal time---$D\delta/D\tau$---at fixed initial overdensity, i.e. $\delta({\mathbf m},\tau_{nl})\equiv Q_\delta(p)$, where
$p$ denotes the percentile in the distribution of initial overdensities at the onset of nonlinearity $\tau_{nl}$,
$Q_\delta(p)$ is the quantile function of that distribution. For explicitness, we are going to use the Dirac delta function
below--denoted here as $\psi$ because $\delta$ is already spoken for---to isolate specific mass parcels that originated from a
set of ``primordial'' ones at $t=t_{nl}$, the epoch at which the relevant fluctuations hosting the seeds of galaxies detached
from the Hubble expansion.

\begin{equation}
\left\langle \frac{D\delta}{D\tau} \right\rangle_{V|p} =
\int_V \int_{-1}^{\infty} \frac{D\delta({\mathbf m},\tau)}{D\tau}
\psi[\delta({\mathbf m},\tau_{nl})-Q_\delta(p)] d \delta({\mathbf m},\tau_{nl}) dV
\end{equation}
where $\psi[\delta({\mathbf m},\tau_{nl})-Q_\delta(p)]$ isolates those overdensities that start out with the same
value at time $\tau=\tau_{nl}$, again, the time when the relevant density fluctuations have decoupled from the Hubble expansion.
Thus,
\begin{equation}
\begin{split}
\left\langle \frac{D\delta}{D\tau} \right\rangle_{V|p} =
-\int_V \int_{-1}^{\infty}
[1+&\delta({\mathbf m},\tau)] [\nabla \cdot {\mathbf u}({\mathbf m},\tau)]\times \\
&\psi[\delta({\mathbf m},\tau_{nl})-Q_\delta(p)] d\delta({\mathbf m},\tau_{nl}) dV
\end{split}
\label{eq:vol}
\end{equation}

Because modes grow independently prior to nonlinearity
\begin{equation}
\begin{split}
\int_V  \int_{-1}^{\infty} [\nabla \cdot {\mathbf u}({\mathbf m},\tau)] \psi[\delta({\mathbf m},\tau_{nl})-Q_\delta(p)]
 d \delta({\mathbf m},\tau_{nl}) & dV \equiv  \\
&\int_V [\nabla \cdot {\mathbf u}({\mathbf m},\tau)] dV
\end{split}
\label{eq:modes}
\end{equation}
rendering the first term in Equation \ref{eq:vol} identically zero due to Gauss's (divergence) theorem. Thus we
are left with
\begin{equation}
\begin{split}
\left\langle \frac{D\delta}{D\tau} \right\rangle_{V|p} =
-\int_V & \int_{-1}^{\infty}
\delta({\mathbf m},\tau) [\nabla \cdot {\mathbf u}({\mathbf m},\tau)] \times \\
&\psi[\delta({\mathbf m},\tau_{nl})-Q_\delta(p)] d \delta({\mathbf m},\tau_{nl}) dV
\end{split}
\end{equation}

Taking the derivative of both sides with respect to $t$,
\begin{equation}
\begin{split}
\left\langle \frac{D^2\delta}{D\tau^2} \right\rangle_{V|p} =
-\frac{D}{D\tau}
\int_V & \int_{-1}^{\infty}
\delta({\mathbf m},\tau) [\nabla \cdot {\mathbf u}({\mathbf m},\tau)] \times \\
&\psi[\delta({\mathbf m},\tau_{nl})-Q_\delta(p)] d \delta({\mathbf m},\tau_{nl}) dV
\label{eq:d2}
\end{split}
\end{equation}
noting that the scale factor $a$ was set at the time of decoupling.
Only one of the terms in Equation \ref{eq:d2} survives integration as nonzero, such that
\begin{equation}
\begin{split}
\left\langle \frac{D^2\delta}{D\tau^2} \right\rangle_{V|p} =
-\int_V \int_{-1}^{\infty}&
\delta({\mathbf m},\tau)
 \biggl[\nabla \cdot \frac{D{\mathbf u}({\mathbf m},\tau)}{D\tau}\biggr] \times \\
&\psi[\delta({\mathbf m},\tau_{nl})-Q_\delta(p)] d \delta({\mathbf m},\tau_{nl}) dV
\end{split}
\end{equation}

Substituting Equation \ref{eq:vel} into the above, one obtains
\begin{equation}
\begin{split}
\left\langle \frac{D^2\delta}{D\tau^2} \right\rangle_{V|p} =
\int_V \int_{-1}^{\infty}
\delta({\mathbf m},\tau)\nabla^2\phi 
\psi[\delta({\mathbf m},\tau_{nl})-Q_\delta(p)] d \delta({\mathbf m},\tau_{nl}) dV
\end{split}
\label{eq:exp2}
\end{equation}


\begin{figure*}
\centerline{
\includegraphics[width=0.30\textwidth]{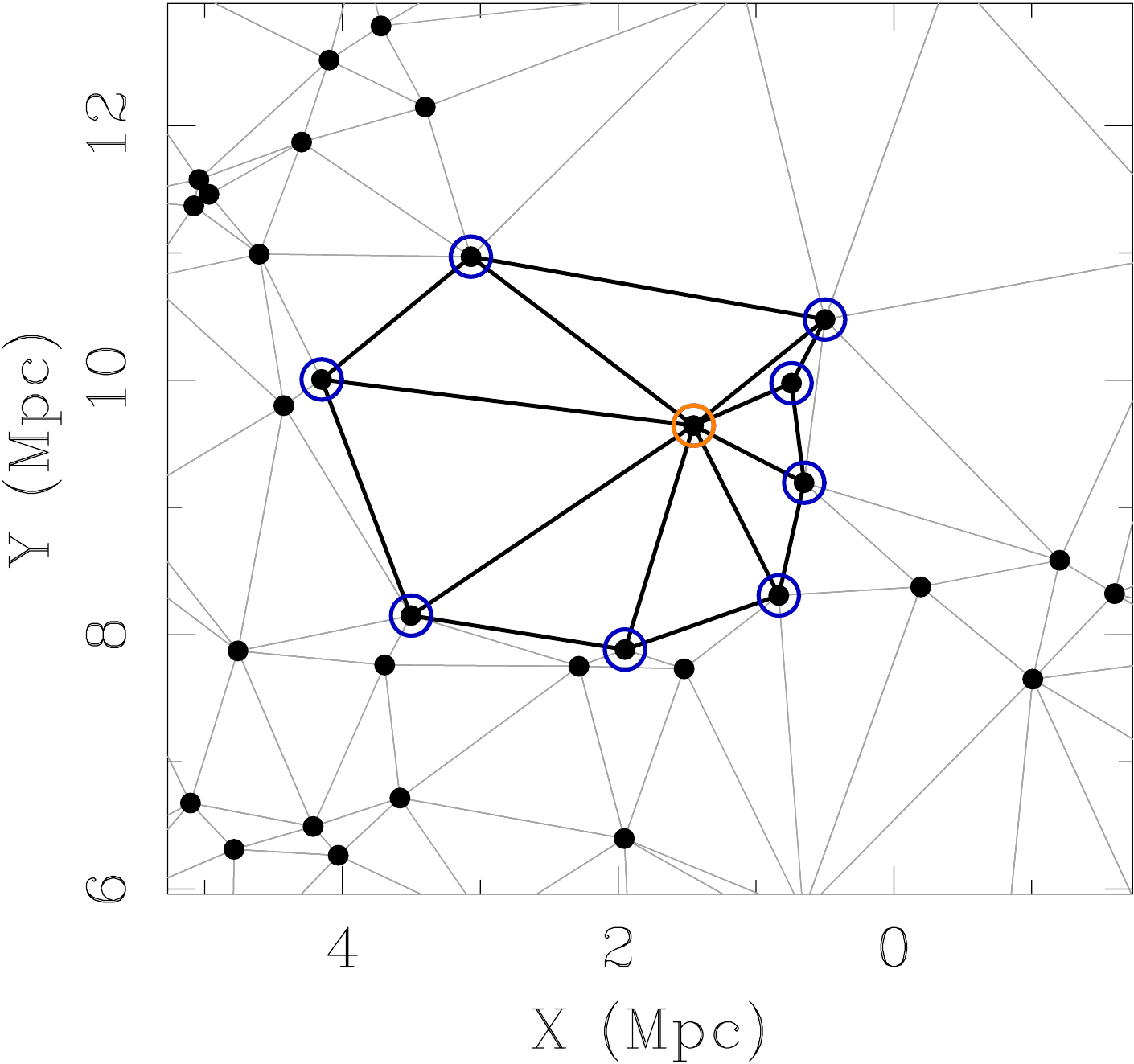}
\includegraphics[width=0.30\textwidth]{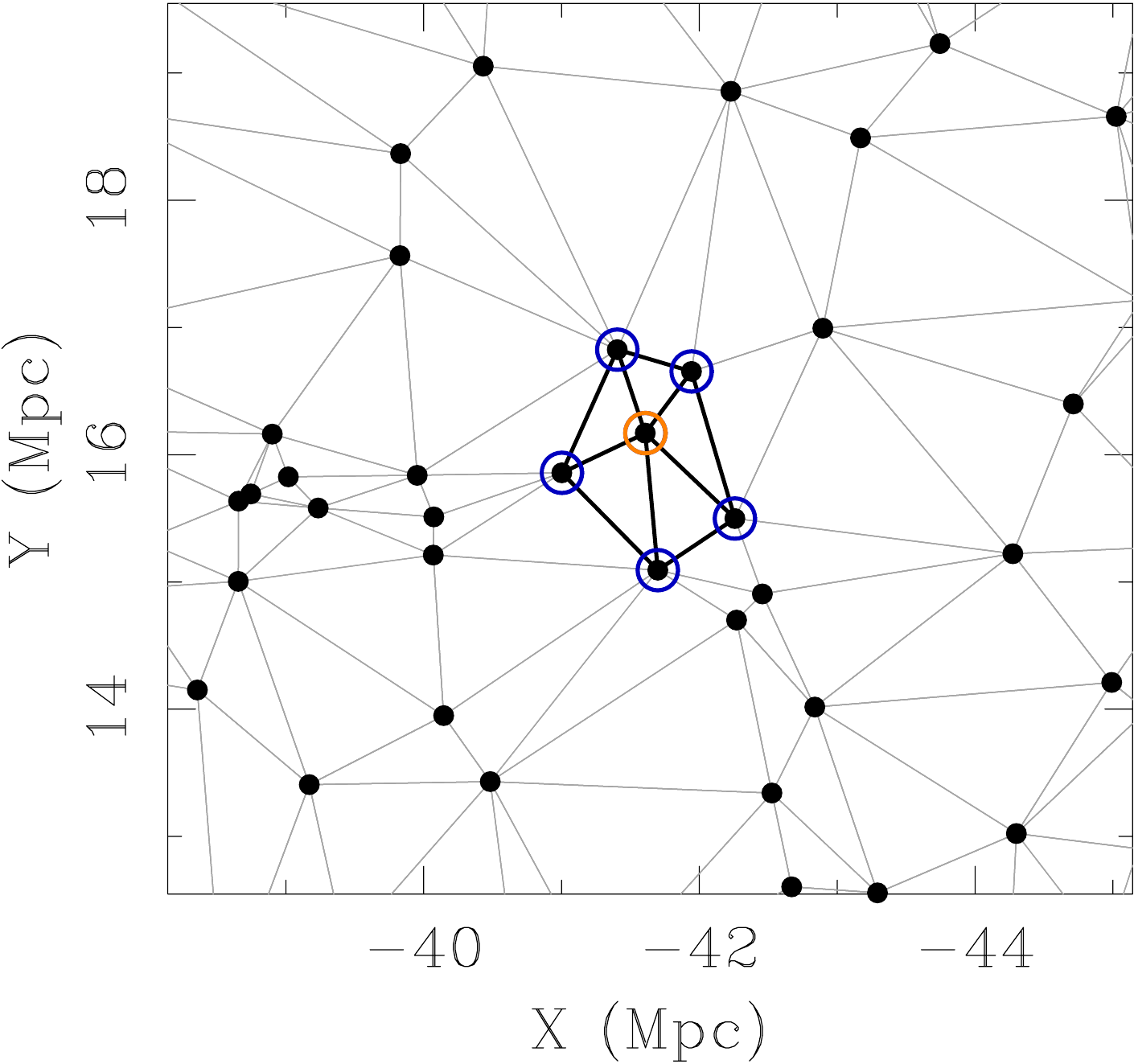}
\includegraphics[width=0.30\textwidth]{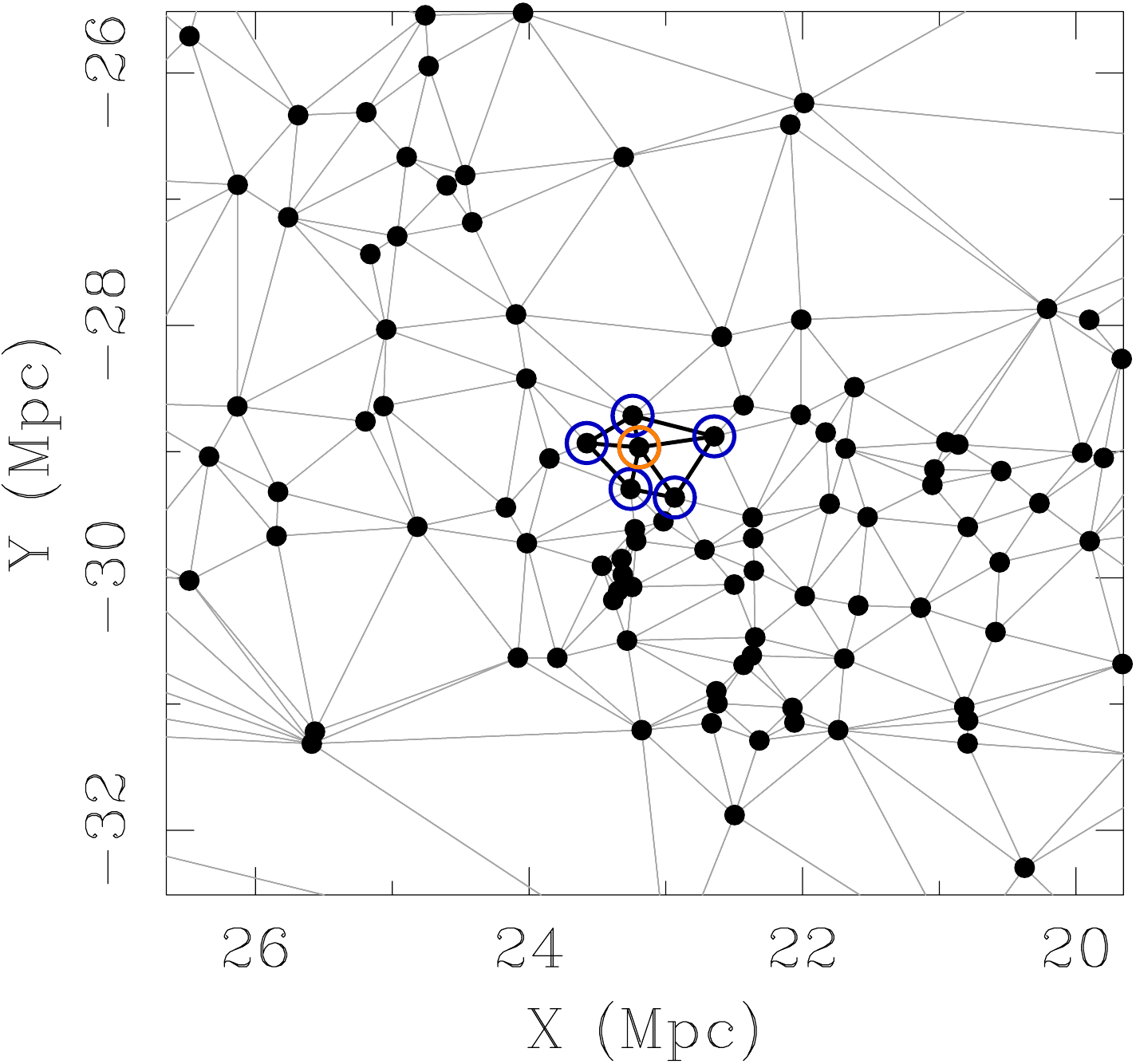}}
\caption{Example Delaunay triangulation of CSI galaxies in the redshift slice $0.48\le z\le 0.60$ within the SWIRE XMM
field. Each box is 7 Mpc $\times$ 7 Mpc. To compute the surface density at the location of a galaxy of interest
(open orange circle) at center of each box, the areas of the adjacent triangles are summed and multipled by the depth of
the redshift slice to derive a local volume element. These volume elements adapt to the point distribution uniquely.
Dividing the completeness-corrected sum of the stellar masses of the galaxies at the vertices of these adjacent
triangles (both blue and orange open circles) by the local volume element, one has constructed an adaptively measured
local stellar mass density. Monte Carlo simulations of incomplete, but correctable, catalogs have verified that these
procedures can accurately recover local densities (see
text for more details)
\label{fig:triangles}}
\end{figure*}


The integration over $\delta({\mathbf m},\tau_{nl})$ isolates those fluctuations that began as
$\delta({\mathbf m},\tau_{nl})= Q_\delta(p)\equiv \delta_p$, so
we can write that the mean second derivative---again at fixed percentile $p$---is simply:
\begin{equation}
\left\langle \frac{D^2\delta}{D\tau ^2} \right\rangle_{V|p} =
4\pi G a^2\rho_M \delta_p^2
\end{equation}
or,
\begin{equation}
\left\langle \frac{D^2\delta}{D\tau ^2} \right\rangle_{V|p} = 
\frac{3}{2}\Omega_M H_0^2 a^{-1} \delta_p^2
\end{equation}

We therefore anticipate that at times $\tau\ge \tau_{nl}$ the ensemble should appear to have a mean growth rate of
\begin{equation}
\left\langle \frac{D\delta}{D\tau} \right\rangle_{V|p} =
\frac{3}{2}\Omega_M H_0^2 a^{-1} \delta_p^2 (\tau-\tau_{nl})
\label{eq:rate}
\end{equation}
The constant of integration is expected to be identically zero over cosmological volumes for modes just detaching
from the Hubble expansion.

Note that the above form is symmetric in $\delta_p$. In other words, parcels with a mass consistent with originating
from the population at $\delta_p=-1/2$ {\it change\/} on average at the same rate as parcels from the $\delta_p=1/2$
population change on average. This is not the same thing as quantifying the amount of matter moved {\it from\/} the ensembles of density
fluctuations defined by $p<0.5$. On average the Universe moves mass out of the population of $p<0.5$ density parcels to
higher density, and we are interested in computing the amount of mass lost from the ensembles of low $\delta_p$in order to properly construct the cosmological evolution in the distribution of densities.

If we define $\delta'\equiv\langle\rho\rangle/\rho -1$, the average amount of mass gained {\it from\/} regions of low density can be
straightforwardly computed, with the chain rule recasting the result in terms of $\delta$, and yielding a result then
valid for the ensembles of parcels with initial densities $\rho<\langle\rho\rangle$ (at least until the budget in material in the $p<0.5$
ensembles has been exhausted).

The two cases for the mean growth/loss trajectories at fixed density percentile are thus:
\begin{equation}
\left\langle \delta \right\rangle_{V|p} =
\begin{cases}
\delta_p +\frac{3}{4}\Omega_M H_0^2 a^{-1}\delta_p^2 (\tau-\tau_{nl})^2 &\text{$\delta_p\ge 0$}\\
\delta_p -\frac{3}{4}\Omega_M H_0^2 a^{-1} \delta_p^2 (1+\delta_p)^2 (\tau-\tau_{nl})^2 &\text{$\delta_p< 0$}
\end{cases}
\label{eq:traj}
\end{equation}
where, again, $\delta_p\equiv \delta(p,\tau_{nl})$, and $\delta_p\ge -1$. When tracked through time, the density at
fixed percentile should therefore follow these trajectories over epochs when sufficient material is available to sustain
growth (or loss)---while none of the individual parcels of matter probably do. As a reminder,
$a$ is set at the epoch of nonlinearity because the density evolution is occuring in regions that have decoupled from
the Hubble expansion at that epoch; i.e., $a\equiv 1/(1+z_{nl})$.

Let us summarize the expectations derived above:
\begin{itemize}
 \item The mean rates of growth for (initial) density fluctuations ought to be larger for regions that
initially had higher density
 \item For density percentiles that will experience unabated collapse---e.g., $p> 0.5$---density growth should scale
in the mean by their initial $\delta^\alpha$, with $\alpha\equiv 2$;
 \item For density percentiles that suffer evacuation---e.g., $p< 0.5$---densities should decline
in the mean by their initial $\delta^\alpha(1+\delta)^2$, again $\alpha\equiv 2$;
 \item The mean trajectory for any density percentiles should be quadratic in time
 $\delta(\tau)-\delta(\tau_{nl}) \propto (\tau-\tau_{nl})^\beta$, with $\beta\equiv 2$, at least while there is
sufficient material to sustain mass flow;
 \item Any additional physics should manifest as additional boundary and/or initial conditions, modifying expectations
 from these canonical values of $\alpha\equiv 2$ and $\beta\equiv 2$
\end{itemize}

We now proceed to an explicit test of whether the evolution of the density distribution indeed follows these
expectations for the growth of structure, employing a large sample of ``stellar mass-selected'' galaxies in CSI, to
directly measure $\alpha$ and $\beta$. For ease, we will revert back to using $t$ in the remainder of the paper as we
work within the framework of the CSI dataset. After studying the evolution of density percentiles in CSI, and what they
mean for the mean growth of nonlinear structure, we discuss a broader set of
implications for future work.


\section{Summary of the Data}
\label{sec:data}

The Carnegie-Spitzer-IMACS Redshift Survey \citep{kelson2014b} was designed to study the relationship between
galaxy growth and environment over the last 9 Gyr of the history of the Universe, the period over which the states and
appearances of today's galaxies were defined and settled. The survey fields were targeted for low-dispersion
spectroscopy with the Inamori Magellan Areal Camera and Spectrograph \citep[IMACS;][]{dressler2011} using, at first, a
three-layer prism described by \cite{coil2011} for the first third of the survey, and then using an innovative
eight-layer disperser for the rest. While both dispersers provide a characteristic resolution of $R\sim 25$,  the newer prism more effectively balances the wavelength
dependence of the spectral resolution, with higher resolution through the red. Again, details can be found in \cite{kelson2014b}. 

CSI chiefly selected galaxies brighter than $[3.6]\le 21$ mag (AB). The effective stellar mass depth for a given
redshift---discussed in more detail by \cite{kelson2014b}---is largely a function of the $M/L$ ratios of stellar
populations, leading to approximate limits of $\sim 10^{10} M_\odot$ for star forming galaxies up to $z\sim 1.5$ and
$\sim 3\times 10^{10}M_\odot$ at $z\sim 1.4$ for galaxies containing old stellar populations. The bulk of the analysis we
perform is with the sample cut at $M_*\ge 10^{10} M_\odot$ when computing local stellar mass densities. When distributions of
densities are computed and analysed, we perform similar analyses using cuts of  $M_*\ge 2\times 10^{10} M_\odot$, $M_*\ge
5\times 10^{10} M_\odot$, and $M_*\ge 6\times 10^{10} M_\odot$ in order to test the robustness of the results.

Once the selection is a few tenths of a dex below $M^*$, our results do not depend sensitively to the detailed choice of
how far down the mass function we compute the stellar mass densities---owing to the fact that the integral of the stellar mass
function is already within $\simlt 20\%$ of convergence. When we (later) normalize local stellar mass densities by the
median density at a given epoch, any systematic change with redshift in the departure from convergence is essentially
removed to first order. And while variations in the shape of the stellar mass function with environment---i.e. with
density---may impose a correlation in the departure from convergence with density at each epoch, this departure itself
will largely evolve at the same rate as the evolution of the stellar mass function in each density percentile. As such, we should
see---to first order---little-to-no systematic error in the density-dependence of the evolution of density at fixed
percentile, and little-to-no systematic error in the time-dependent term of the evolution of density at fixed percentile.
Furthermore, given the magnitude of the difference in $M^*$ between the field and the richest clusters, $< 0.1$ dex at
$z\sim 1$ \citep{vdb2013}, and the fact that such rich environments make up a tiny fraction of the volume of the
universe, we expect such systematic errors to be significantly smaller than our formal/statistical errors over most of
the density range being probed.

Had we counted local galaxy number densities, then these systematics would (a) be much larger, since the mass function
converges very slowly in galaxy number density, and (b) the differential selection between star forming and quiescent
galaxies at low stellar masses could leave a larger imprint as differential growth rates between low and high density percentiles.
Furthermore, any dependencies of the rates of galaxy-galaxy merging on environment would also bias the results, whereas
stellar mass is preserved under merging. In principle a comparison of number and stellar mass density evolution could
provide insight into galaxy merger rates when differential selection is carefully modeled, and we leave such work for a
another day.

In the SWIRE XMM field, we use a high quality set of $51\,001$ observations of $42\,251$ galaxies, and in
the SWIRE CDFS field, a high quality set of $40\,680$ observations of $35\,867$ galaxies. Duplicate
observations remain in our analyses, with each observation being assigned a reduced weight of $1/N_{obs}$ so
that the sum of the weights for a given galaxy observed more than once would equal the weight for singly
observed galaxies.

These samples have been reanalysed since the publication of \cite{kelson2014b}, owing to
a growing understanding that the diversity of galaxy formation histories is not only real
\citep[e.g.][]{abramson2016}, but mathematically constrained \citep[e.g.][]{kelson2014,kelson2016,dressler2018}. Thus,
we undertook an extensive effort to refit the CSI SEDs using a library of star formation histories
generated in manner that reproduces the observed distributions of specific star formation rates of real galaxies
\citep{kelson2014}.


\begin{figure}
\centering
\includegraphics[width=0.40\textwidth]{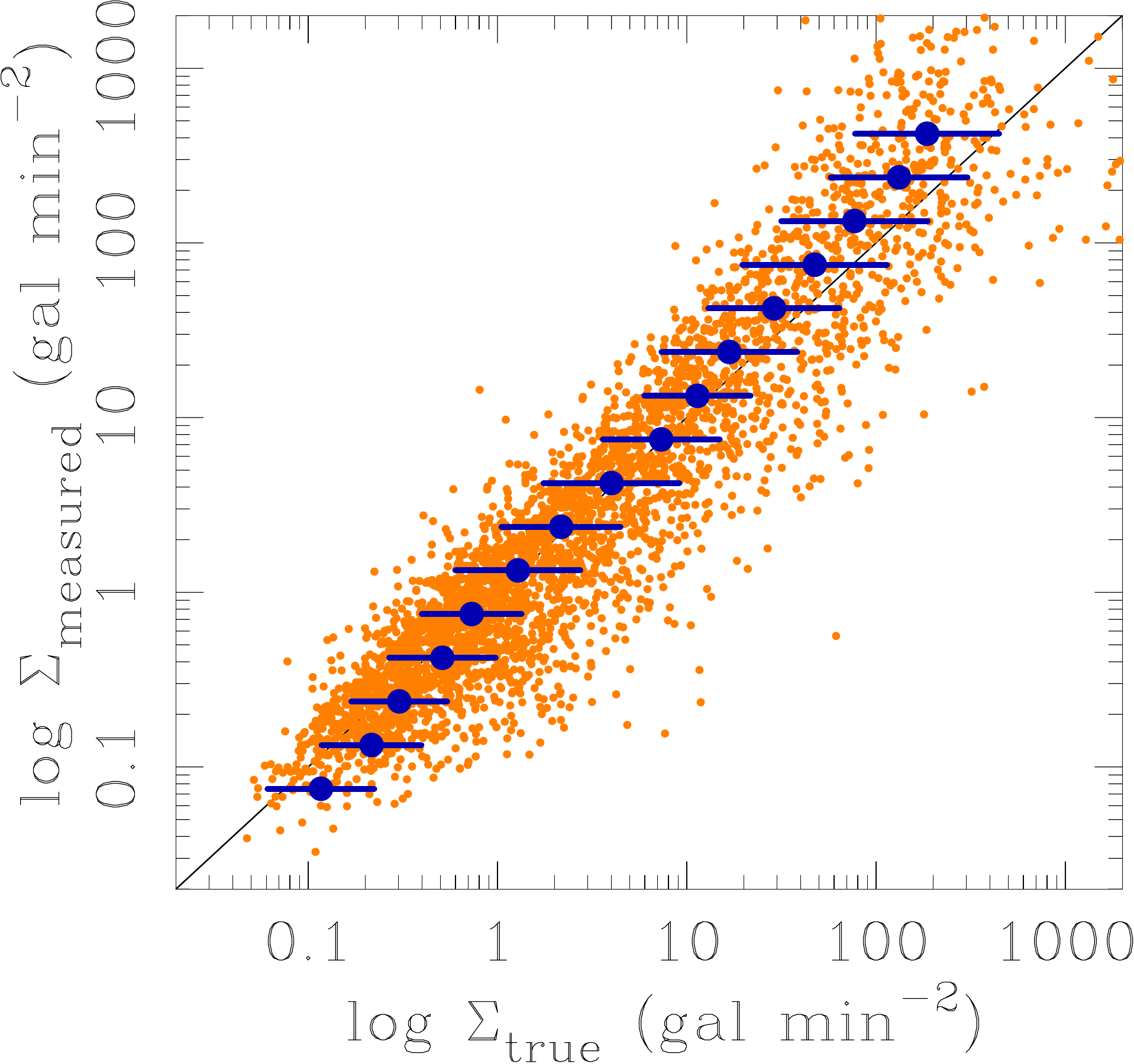}
\caption{The results of Monte Carlo simulations of distributions of galaxies
in fields of variable mean projected density on the sky, in which {\it local\/}
projected densities are computed using the Delaunay Triangulation estimator
described in the text. Here we plot local projected density estimates from
catalogs suffering from source-density-dependent incompleteness in a manner
similar to the CSI selection function, plotted against what the local projected
densities would have been without such incompleteness. The general correspondance
along the unity line indicates that the effects of slit collisions or other
source-density-dependent sources of incompleteness do not systematically bias
our density estimators (red points). In blue we bin the values of
the true projected surface densities---showing the median and robust standard deviation of $\Sigma_{true}$---
at fixed $\Sigma_{measured}$, confirming that survey incompleteness can be reliably
corrected to recover the distributions of true projected densities.
\label{fig:demo}}
\end{figure}


Because the histories underpinning the SED fitting are better constrained, span properties that are more
realistic---by definition---and have significantly fewer free parameters, we were able to recover redshifts
for more objects and to fainter optical magnitudes, with an average completeness of $\sim 50\%$ over $19\
\hbox{mag} \simlt i_{AB}\simlt 23$ mag, declining to $\sim 30\%$ at $i_{AB} \sim 24$ mag and $\sim 25\%$ at
$i_{AB} \sim 24.7$ mag. As described in \cite{kelson2014b}, corrections for incompleteness are trivariate
functions of magnitude, colour,  as well as the local density of sources in the original IRAC
catalog.\footnote{While the completeness corrections used here rely on source-densities computed in boxes with 0\Min 5
sides, we have verified that our results do not depend on this choice. This insensitivity is due to the fact that
overdensities in redshift space do not correspond well to overdensities in the IRAC target catalog except at the
locations of the rarest of very rich/massive groups.}

In comparisons with previously published ``high-resolution'' redshifts \citep{lefevre2003,cooper2012,scodeggio2018},
we find that over the ranges of stellar mass and redshifts
we are investigating in this paper---$M\ge 10^{10}M_\odot$---the CSI redshifts have typical errors of
$\Delta z\approx 0.01(1+z)$ for galaxies at $0.5\le z\le 1.0$, and $\Delta z\approx 0.02(1+z)$ for galaxies at
$1.0\le z\le 1.4$. The rate of catastrophic outlyers ($|\Delta z/(1+z)|>0.15$) is of order $2-3\%$---from a
mixture of our own catastrophic failures and those in published redshift catalogs themselves. This level of accuracy is
adequate for the analysis we propose, in which we measure local stellar mass densities in fairly wide redshift
slices.

More details on the derived properties of the sample will be used in future analyses of the galaxy
populations, but the work presented here relies solely on the most robust outputs of the SED fitting:
redshifts and stellar masses.


\begin{figure}
\centering
\includegraphics[width=0.40\textwidth]{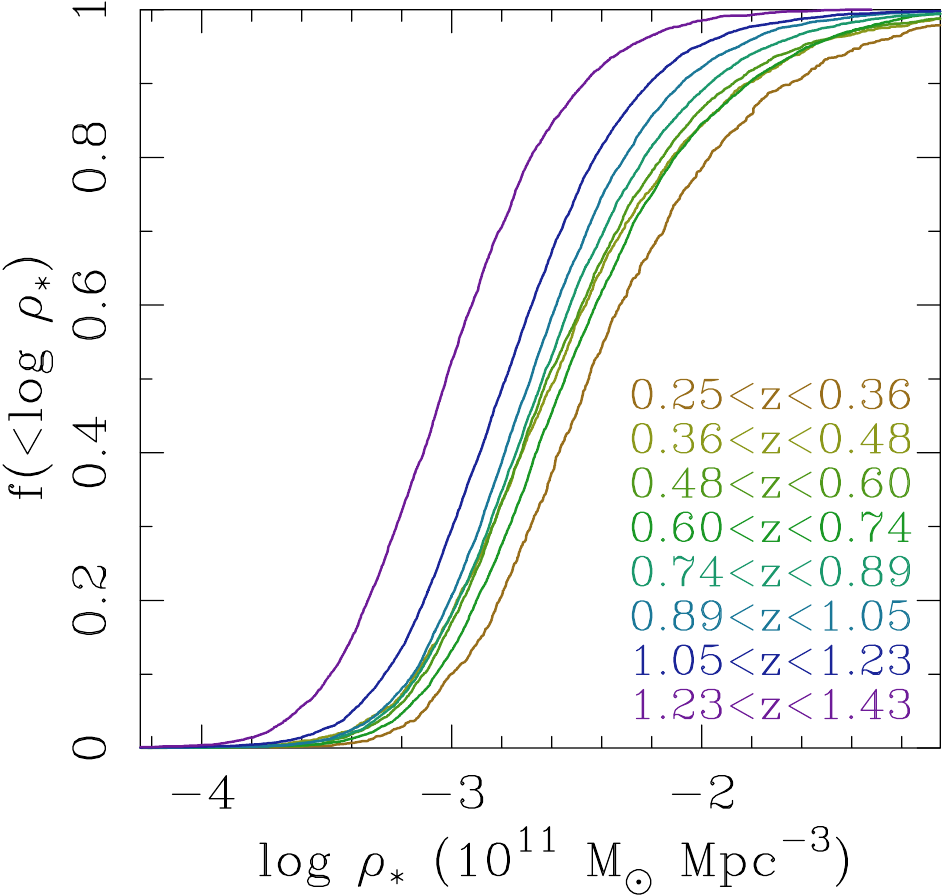}
\caption{The cumulative distributions of local stellar mass densities from the combined samples of the
CSI SWIRE XMM and CDFS fields in eight redshift slices from $z=0.25$ to $z=1.4$.
\label{fig:CumulativeDist}}
\end{figure}



\section{Measuring Local Volume Densities in CSI}
\label{sec:local}

Here we discuss our procedure for converting galaxy point data to a map of densities, but one that is not referenced to
a specific physical scale. Outside of the nonlinear regime the evolving density field is well-understood through its
decomposition into Fourier modes in an Eulerian frame of reference. Typically such scales are investigated using
densities measured at specific scales or wavenumbers, and counting galaxy point data in such cells has a long tradition of
being used to probe the underlying distribution of matter since at least \cite{coles1991}. But our expectations for the
evolving distributions of local densities were derived specifically in a Lagrangian coordinate system, and, as such, we
require densities not measured over a specific scale but be adaptive to the spatially variable structures in the galaxy
distributions.


\begin{figure*}
\centering
\includegraphics[width=1.00\textwidth]{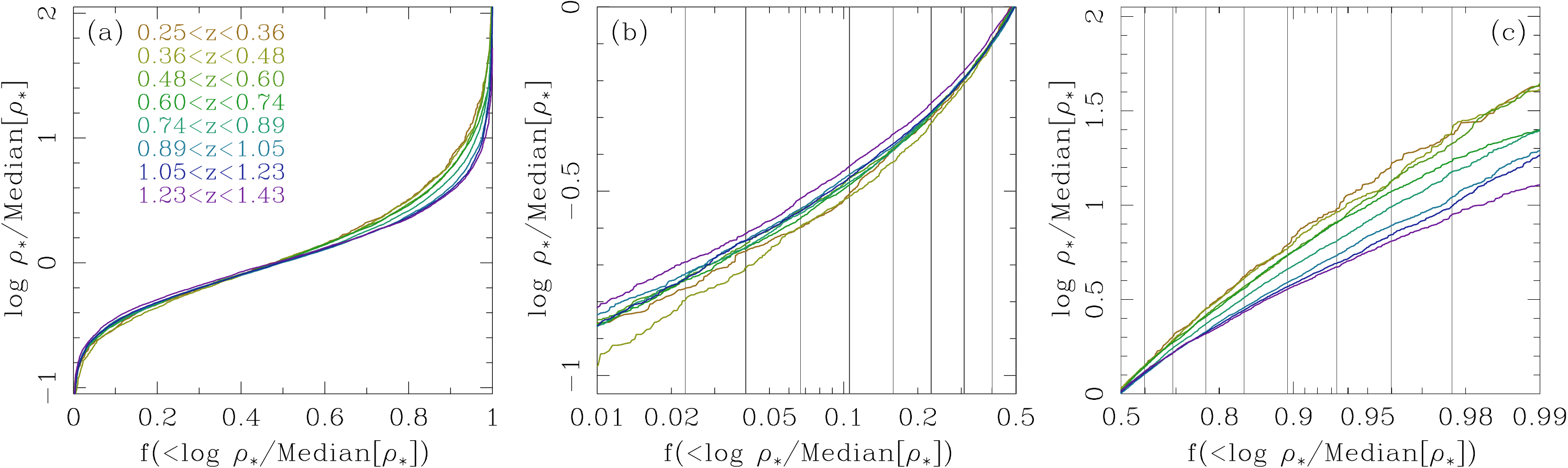}
\caption{(a) The cumulative distributions normalized to the median density at each redshift.
(b) Magnification of the low density tail of the cumulative distributions, with vertical lines at
the percentiles displayed in Figure \ref{fig:DensityDist}.
(c) Same as in (b) but for the high densty tail.
\label{fig:EvolDist}}
\end{figure*}


Adaptive measures of local densities have been used previously to better understand how and why galaxy properties
are correlated with environment, such as \citet{dressler1980}'s use of the radius enclosing the $10$th-nearest neighbour
to define a local element of area.\footnote{It is also interesting to note that this work presciently identified
the importance of early conditions in setting up the long-term divergence in evolutionary states for galaxies
in different late-time environments.} We opt here for a more adaptive---and computationally simple---approach, one that
is not only fast but easily made robust against sources of survey incompleteness such as that arising from slit
collisions or luminosity selection.

Briefly, for a list of objects in a fixed redshift slice, we run a Delaunay triangulation \citep[part of the
Visualization Toolkit;][]{VTK}. This procedure creates a unique list of triangles with galaxies at each of the
vertices, but maximizes the angles to avoid particularly thin triangles.
The effect of this optimization is that when circles are drawn to
circumscribe the triangles, those circles contain no other points.
We show three examples from the Delaunay triangulation from one redshift slice, $0.48\le z\le 0.60$ in the SWIRE
XMM field of CSI in Figure \ref{fig:triangles}.
These examples span a range of local densities as can be seen from the distributions of the
galaxies (black filled circles) in each 7 Mpc $\times$ 7 Mpc box, and the lighter line segments outlining
the triangles.

We then compute local projected area elements at the location of a given galaxy (e.g. the galaxies circled in orange)
by summing the areas of the triangles (shown in solid black) connected to that galaxy. The sizes of the triangles scale
directly with the local separations of galaxies; in regions of high number density these triangles shrink
commensurately. When the sum of the triangle areas is multiplied by the comoving length of the redshift slice, we have a
local volume element. If we had a complete catalog containing all galaxies, we would define the density as the number of
unique vertices (i.e. number of orange $+$ blue open circles) divided by that local volume element. A stellar mass
density would be defined as the sum of the stellar masses of the galaxies at these vertices, divided by the local volume
element.

However, no redshift survey is 100\% complete and this procedure must be corrected for incompleteness. To do so we
incorporate galaxy weights: each galaxy thus represents ones just like it in colour, magnitude, and local source-density
but are missing from the survey. The weights are defined as the inverse of the completeness for a given galaxy's colour,
magnitude, and local source-density.\footnote{This approach was validated in \citet{kelson2014} by verifying that
completeness-corrected local counts systematically reproduce that of the original photomety catalog.}
Completeness-corrected local densities are thus computed by summing the weights of
the galaxies and then dividing by the volume element.\footnote{The distributions of local volume densities we obtain are
insensitive to variations in how we define the source-density dependent term of our completeness function(s).}
Completeness-corrected stellar mass densities are computed by summing the products of the weights and stellar masses
of the galaxies, with subsequent division by the volume element.

As a reminder, we focus this paper on stellar mass densities because (1) the steepness of galaxy luminosity and stellar
mass functions leads total number counts to converge slowly, whereas the integral of stellar mass converges much more
rapidly, and (2) number counts are not preserved under galaxy-galaxy merging whereas stellar mass remains conserved
(modulo tidal ejection or stripping of loosely bound stars).

In redshift surveys such as CSI, slit collisions reduce the efficiency with which targets can be observed in
regions of high source density. We demonstrate the reliability with which we could recover local
number density fluctuations using Monte Carlo simulations of CSI-like survey sampling. By generating
fields of galaxies with the mean CSI source density on the sky, and adding a broad range of overdensities 
of varying amplitudes and sizes, and then ``observing'' them with similar source-density dependent incompleteness to
CSI, we measured local densities using our procedures and compared them to the true local densities that
would have been derived from a complete and unbiased catalog.

The results of our tests are shown in Figure \ref{fig:demo}, in which the
measurements of local density in the original, unculled catalog are plotted along the x-axis,
while the incompleteness-corrected local number densities are plotted along the y-axis.
We conclude from these simulations that so long as we measure local projected densities in
CSI using redshift slices that are larger than the typical redshift uncertainties, our
incompleteness-corrected local density measurements should be robust.

Figure \ref{fig:XMMslices} illustrates the spatial distribution of galaxies with stellar masses $M_*\ge 10^{10}$
M$_\odot$ and their local {\it stellar mass\/} densities within four redshift slices of $\sim 5.25$ square degrees of
CSI data. The galaxy points are colour coded by local density relative to the median in each redshift slice.

The expectation---as derived in \S \ref{sec:frame}---is for high density regions to grow in density contrast
relative to the median over time, and for the low density regions to become increasingly hollowed out. And while
individual over- and under-densities cannot be tracked with lookback time, the expectation for the average behaviour
of over- and under-densities can be seen visually in these data.

Qualitatively one sees a striking increase in density contrast towards late times, as the topology of the spatial
distributions of the galaxies evolves---to tighter filaments and knots at later times, as well as to increasingly larger
voids. Note that we do not address such evolution in the specific topological features of the galaxy distributions, but
only here discuss the quantitative evolution in the global distributions of local densities.

To construct cumulative distributions of local density, we simply reorder the arrays of local densities and galaxy
weights in increasing order of density, with a subsequent cumulative summation of
the weight array. Normalizing the resulting array by the sum of these weights yields an array easily sifted
through to identify the array positions of density percentiles. Confidence intervals are estimated
with a distributional method \citep[see][and references therein]{statint} in which the probabilities
that data points may be higher or lower than a given percentile are calculated using the binomial distribution.

Figure \ref{fig:CumulativeDist} shows the cumulative distributions of local stellar mass densities in eight
redshift slices from $z=0.25$ to $z=1.4$. A bulk increase in stellar mass density can be seen as the distributions
shift towards higher mean/median values of stellar mass densities. However, for the purposes of using the distributions
of local densities to characterise the nonlinear growth of structure, we want to study the distributions of
normalized densities---normalized, that is, by the median in each redshift slice. These are shown in
Figure \ref{fig:EvolDist}(a), in which the distributions look qualitatively similar to each other, but are
indeed distinct in their tails, as magnified in
Figure \ref{fig:EvolDist}(b) and (c).

Figure \ref{fig:EvolDist}(b) magnifies the low density tail while Figure \ref{fig:EvolDist}(c) magnifies
the high density tail. In these plots one can directly see that from early- to late-times the high density tails evolve
to increasingly larger density contrasts, and in a manner that is not identical at each density percentile.
Simultaneously, the regions of low density evolve to even lower density contrasts, also by amounts that vary with
percentile. In the next sections we interrogate these distributions, to determine whether these increases and decreases
of density contrast with time, and their dependence on density, are
consistent with the analytical forms derived in \S \ref{sec:frame}.


\section{The Observed Distributions of Local Densities}
\label{sec:evolution}

\begin{figure}
\centering
\includegraphics[width=0.45\textwidth]{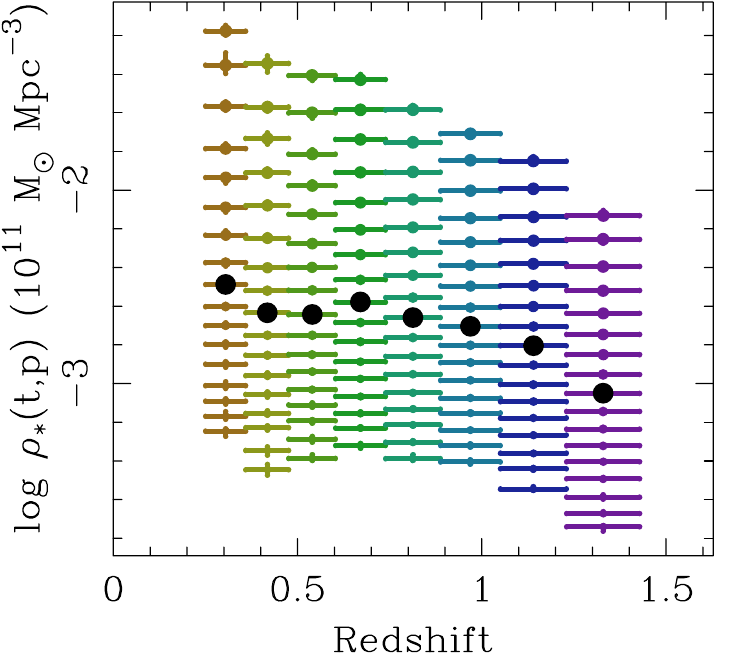}
\caption{The distribution of local stellar mass densities from the combined samples of the
CSI SWIRE XMM and CDFS fields. Each point represents a fixed percentile, $p$, defined
over the range $-2\sigma$ to $+2\sigma$ for a Gaussian distribution, in intervals of $\sigma/4$.
The black filled circles mark the median (i.e. 50th percentile) at each redshift. Cosmic variance
is expected to contribute to the scatter from one redshift to the next at a level of $\sim 0.15$ dex at
$z\sim 0.3$ and $\sim 0.05$ dex at $z\sim 1$, depending on the width of the redshift slice we use
\citep[calculated using mock catalogs from the MDPL2 simulations][]{klypin2016,knebe2018}.
\label{fig:DensityDist}}
\end{figure}

Using the local stellar mass densities, $\rho_*$, measured at the locations of the galaxies in CSI, we now
investigate how the density distribution evolves with cosmic time.
Figure \ref{fig:DensityDist} plots local stellar mass densities for 17 percentiles in the overall
density distribution in eight redshift slices from $z\sim 0.3$ to $z\sim 1.4$, with the percentiles,
$p$, defined as equivalent to the $-2\sigma$ to $+2\sigma$ for a Gaussian distribution, in intervals
of $\sigma/4$. The large black circles mark the median (50th percentile) at each redshift. Cosmic
variance is a significant contributor to fluctuations in the distribution with redshift. Using the MDPL2 simulation catalogs
of mock galaxies \citep{klypin2016,knebe2018} to estimate the magnitude of cosmic variance for CSI-like
volumes, we find that the median density can fluctuate by $\sim 0.15$ dex at $z\sim 0.3$ and $\sim 0.05$ dex at $z\sim 1$.

Figure \ref{fig:DensityDist2} shows these distributions of densities relative to the median in each redshift slice. When
the density percentiles are normalized relative to the median, the fluctuations due to cosmic variance are greatly
diminished and typically less than 0.05 dex, with the exception of the highest percentiles ($p\simgt 0.95$) in the two
lowest redshift slices shown here.

With these data alone one cannot necessarily say how individual density fluctuations grow with time in the nonlinear
regime, but with these data we can measure mean rates of evolution for each percentile. As is visible in the figure,
higher percentiles have densities that grow in the mean more rapidly with time than lower percentile overdensities. 
Underdense regions, i.e. traced by percentiles $p<0.5$, appear to show negative ``growth'' in the mean. Together these
data will be shown below to greatly constrain the nature of the stochastic process that is the nonlinear growth of
structure.

For each aspect of our analysis of the CSI redshift catalogs we have varied the widths of the redshift
slices from 200 comoving Mpc to 400 comoving Mpc and find that our results remain robust, with the basic
pattern shown in Figures \ref{fig:DensityDist} and \ref{fig:DensityDist2}. We have also performed the same
analysis on catalogs of galaxies in the MDPL2 simulations \citep{klypin2016,knebe2018} and find similar patterns.


\section{Making Inferences from the Evolution of the Density Percentiles}
\label{sec:info}

\begin{figure}
\centering
\includegraphics[width=0.45\textwidth]{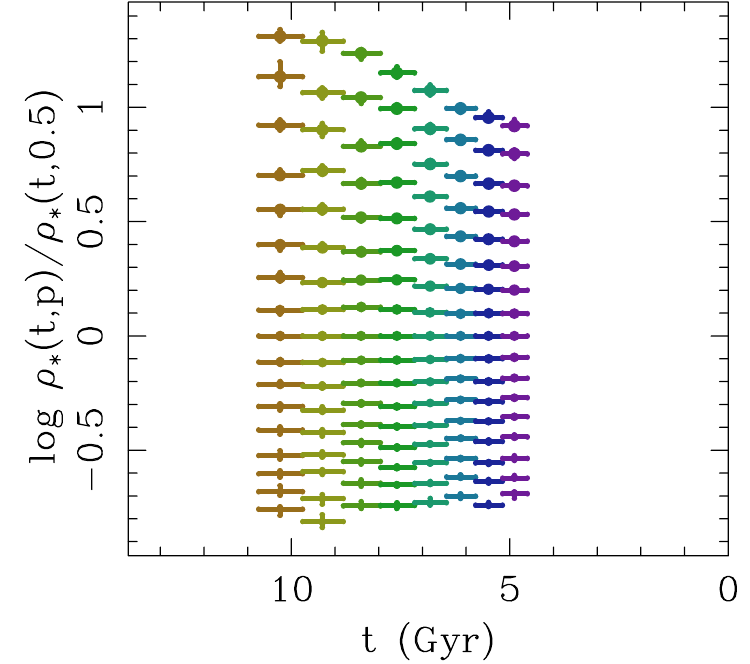}
\caption{The distribution of local stellar mass densities relative to the median within a redshift
slice. Relative to the median local stellar mass density, higher density percentiles are growing
more rapidly than lower density percentiles.
\label{fig:DensityDist2}}
\end{figure}

In \S \ref{sec:frame} we derived a set of expectations for how the local mass density attached to a given
percentile in the density distribution should evolve: quadratic in time, with a quadratic
dependence on the initial density for $p>0.5$, and a quartic dependence on initial density for $p<0.5$.
If one knows {\it a priori\/} the distribution of density fluctuations transitioning to nonlinear growth
that are traced by galaxies at the start of star formation, then one has the required mapping of $p\rightarrow
\delta_p$ to begin modeling the CSI observations in Figure \ref{fig:DensityDist2}.


While we expect that distribution of densities to follow something like a lognormal distribution
\citep[e.g.][]{coles1991,wang2011} with a standard deviation of order unity, we first employ the CSI data to test and confirm
 these expectations---by evolving the individual density percentiles back in time to the epochs when star formation may have
began (i.e. $t=t_0$ at $z\simgt 10$). Once the form of the initial density distribution is understood, we will then use
that functional form---with any additional and required nuisance parameters to be marginalized over---in fitting a
reasonable set of of percentiles simultaneously. In this section we treat and fit each percentile independently and
interpret what those individual fits yield.

For a given percentile, $p$, let us write that $\rho_*(t,p)/\rho_*(t,0.5)$ evolves according to
\begin{equation}
\frac{\rho_*(t,p)}{\rho_*(t,0.5)} = A(p) + B(p)(t-t_0)^\beta
\label{eq:rho}
\end{equation}
where $t$ is the time since the Big Bang and $t_0$ is the age of the universe when stellar mass
started to form. Note that we have switched from $\tau$ in \S \ref{sec:frame} to the more familiar $t$
for cosmic time.
By definition $A(p)$ is the $p$th percentile density at the start of stellar mass
growth, and $B(p)$ is the mean rate of growth for the ensemble of those initial densities.
For this exercise we treat the $A$'s and $B$'s as independent. Once we know the functional form of
the initial conditions, we will be able to abstract and generalize Equation \ref{eq:rho}.

For the purposes of the figures, we adopted a power-law evolution in time with $\beta=2$ but note that similar
results are found for the inferred initial density distribution had we adopted any value at least over $1\le
\beta\le 3$---though slightly different values for the standard deviation, $\sigma$, of the lognormal would
have resulted.


\begin{figure}
\centering
\includegraphics[width=0.45\textwidth]{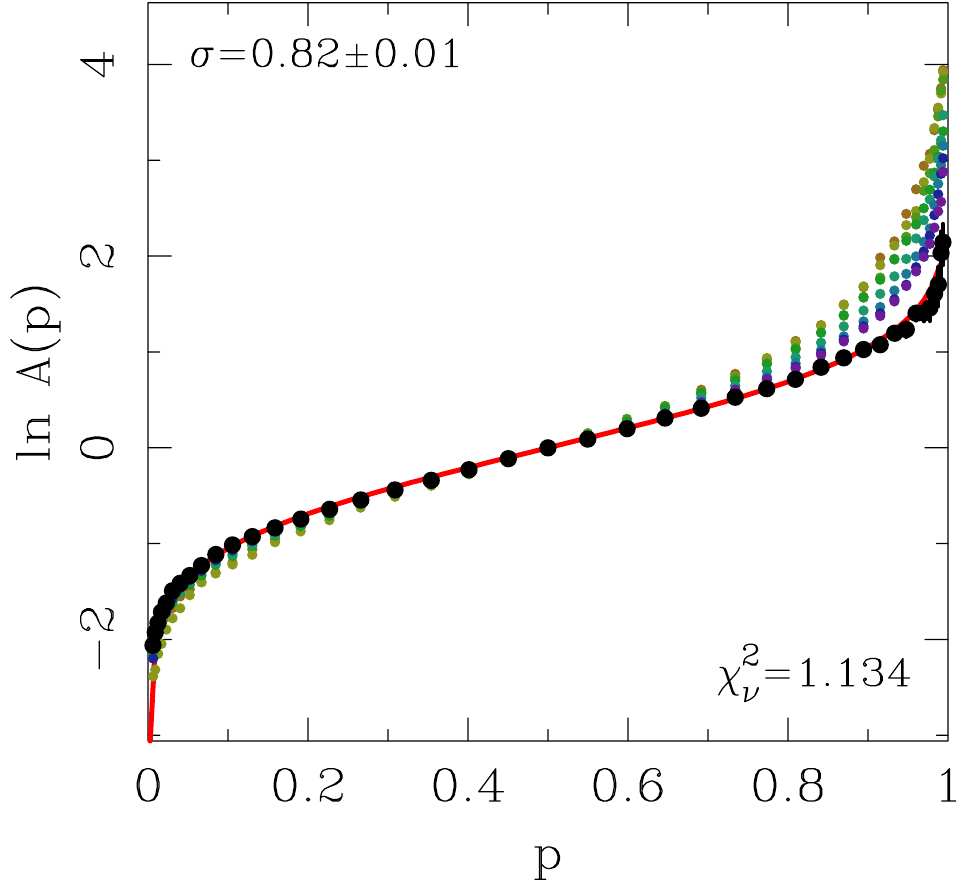}
\caption{The estimated initial densities, $A(p)$, for each percentile, are shown with the black
points, as derived from fitting the evolution of the densities for each percentile as separate
power-law functions of time as described by Equation \ref{eq:rho}. Individual measurements of
relative density from seven redshift slices are using the coloured points. A least-squares fit
for the lognormal distribution of $A$ that best fits the black points was performed, shown by the
red line, with $\sigma=0.82\pm 0.01$. Evolving the highly non-lognormal distribution
of densities back in time to the start of galaxy growth yields a lognormal distribution of early
densities to a high degree of precision and within the measurement errors, independent of our
redshift binning, and independent of the assumption that $\beta=2$.
\label{fig:Lognormal}}
\end{figure}


Assuming $\beta=2$, we derive a correlation of $A(p)$ with percentile, $p$, shown by the black points in
Figure \ref{fig:Lognormal}. For comparison with the individual epochs in CSI, data points for
$\rho_*(t,p)/\rho_*(t,0.5)$ are also shown using the smaller coloured points (these were derived from an analysis
using seven redshift slices over the same epoch, to reduce clutter). At late times, the
distribution of densities is, as is well known, far from lognormality. But when we extrapolate the power-law evolution
of each density percentile back to $t=t_{nl}$, we obtain a distribution of densities that is consistent with being lognormal. 
We performed a least-squares fit for the lognormal distribution that best represents $A(p)$, inflating
the formal errors in $A(p)$ by $\sim 2.5\%$ in quadrature to account for cosmic variance (expected to be $\sim 1/10$th the
redshift-to-redshift variation in the median density).
To within our observational errors---reducecd $\chi^2=1.134$ per degree of freedom---we
find that a lognormal provides an accurate description of the density distribution inferred for the epochs
when star formation began, with $\sigma=0.82\pm 0.01$ (shown by the solid red line). Restated in the context
of \S \ref{sec:frame}, $A(p)\equiv Q_\delta(p)+1\equiv \exp{[\sigma\Phi^{-1}(p)]}$, where $\Phi^{-1}(p)$ is the probit
function\footnote{The probit function is the inverse of the normal cumulative distribution function.} and $\sigma$ is
the standard deviation of the initial lognormal density distribution.


\begin{figure}
\centering
\includegraphics[width=0.42\textwidth]{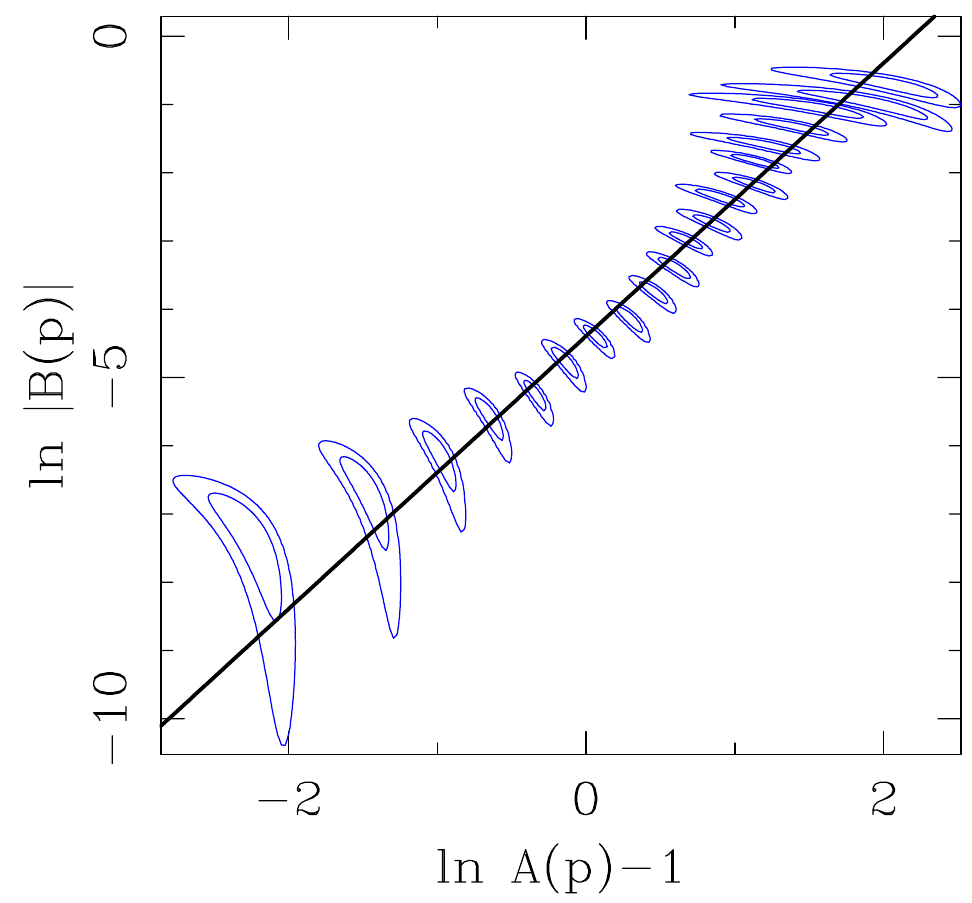}
\caption{The correlation of $B(p)$ with $A(p)-1$ for $p\ge 0.5$, as derived from
fitting the mean change in density at fixed density percentile with the power-law
form of Equation \ref{eq:rho}, assuming $\beta=2$, and noting that that $\delta_p\equiv A(p)-1$. The
expectation of $B(p)= \frac{3}{4}\Omega_M H_0^2 (1+z_{nl}) [A(p)-1]^2$ is shown by the black line (assuming
$z_{nl}=10$). While the inferred rates of growth appear consistent with a logarithmic slope of $\alpha=2$, we fit the
density- and time-dependence of the mean evolution of structure in the next section to simultaneously derive confidence
intervals on $\alpha$ and $\beta$.
\label{fig:Gravity}}
\end{figure}


In Figure \ref{fig:Gravity} we plot the measured values of $B(p)$ for positive overdensities---$p> 0.5$---versus the corresponding values of
 $A(p)$ from the fits to the evolution of the individual percentiles. Comparing Equation \ref{eq:rho} to the expressions derived
in \S \ref{sec:frame} we identify $\delta_p\equiv A(p)-1$, and thus plot the expectation $B(p)\propto [A(p)-1]^2$
by the solid black line, including the expected normalization. Note that, in general, departures in the
normalization will be sensitive to the sizes of the redshift slices in our analysis and, in particular, how their sizes
compare to the typical redshift uncertainties, among other systematic effects. As was clearly visible in Figure
\ref{fig:DensityDist2}, regions of high density grow in the mean at rates
significantly greater than regions of low density, and Figure \ref{fig:Gravity} indicates that the data appear
consistent with expectations of $\alpha=2$.

\begin{figure}
\centering
\includegraphics[width=0.42\textwidth]{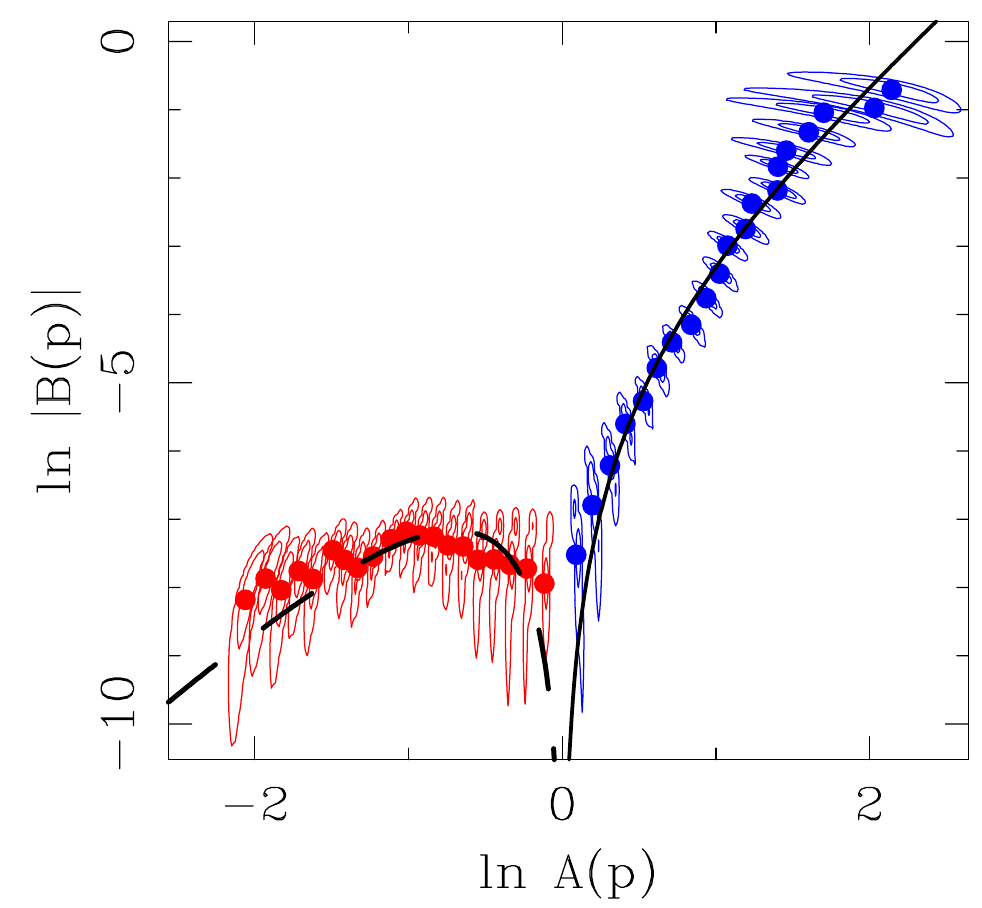}
\caption{Same as in Figure \ref{fig:Gravity} but now including the low density percentiles in red.
The high density expectation of $B(p)= \frac{3}{4}\Omega_M H_0^2 (1+z_{nl}) [A(p)-1]^2$ is shown by the solid black line,
while the low density expectation of $B(p)= -\frac{3}{4}\Omega_M H_0^2 (1+z_{nl}) [A(p)-1]^2 A(p)^2$ is shown using the
dashed black line. While accurate measurements at low density are more hampered by systematic uncertainties in the
analysis of the CSI data set compared to high density regions, the predictions and mean observed
evolution for regions losing mass are in qualitative agreement. In the next section, when we fit for the density- and
time-dependence of the mean evolution of structure simultaneously, we include only measurements at $\Phi^{-1}(p)\ge -1$
to mitigate the larger effects of systematic uncertainties at low density.
\label{fig:Unified}}
\end{figure}

The accurate measurement of evolution in the low density percentiles can be hampered by incompleteness, as well as
large and/or variable redshift errors, at least releative to the sizes of the redshift slices. Nonetheless, we plot
in Figure \ref{fig:Unified} both the low and high density regimes from the data binning used in Figures
\ref{fig:DensityDist} and \ref{fig:DensityDist2}. The inferred values of $A(p)$ and $B(p)$ for $p<0.5$ are
shown in red, while blue is retained for $p> 0.5$. As a reminder, the predicted dependence in the low density regime
is $B(p)\propto [A(p)-1]^2 A(p)^2$, now shown by the dashed black line. The solid black line, as in Figure
\ref{fig:Gravity}, shows the prediction for positive overdensities. The measured evolution of both high and low density
percentiles appear qualitatively consistent with the expected density-dependencies derived in \S \ref{sec:frame}.

We now proceed in the next section to jointly infer confidence limits on both $\alpha$ and $\beta$, adopting a lognormal form
for the density distribution at the start of stellar mass growth with an unknown $\sigma$. Any quantitative constraints
on $\alpha$ and $\beta$ will give us insight into the nature of the process that is the growth of structure.


\section{The Mean Trajectories of Nonlinear Growth and Their Dependencies on Initial Density and Cosmic Time}
\label{sec:global}

In the previous section we showed that the individual density percentiles at late times can be monotonically
related to equivalent percentiles in an initial lognormal spectrum of densities. We use this information now to
construct a simple parameteric model for the mean growth of all the positive overdensities of the
form shown in Equation \ref{eq:rho}, adopting
\begin{eqnarray}
&A(p)=\exp{[\sigma\Phi^{-1}(p)]}\\
&B(p)=\begin{cases}
+\gamma C [A(p)-1]^\alpha (t-t_0)^\beta & p\ge 0.5\\
-\gamma C [A(p)-1]^\alpha A(p)^2 (t-t_0)^\beta & p<0.5
\end{cases}
\end{eqnarray}
where $\Phi^{-1}(p)$, again, is the probit function and $\sigma$ is the standard deviation of the initial lognormal
density distribution (with $\sigma$ now treated as an unknown). Our derivations in \S \ref{sec:frame} predicted that
these histories should be normalized by $C\equiv \frac{3}{4}\Omega_M H_0^2 (1+z_{nl})$, but we include $\gamma$ as an
of-order-unity factor in the fit to account for (a) systematics that can dilute the observed densities
relative to the median; and (b) a systematic departure of the mean epoch of nonlinearity from
our adoption of $z_{nl}=10$. While the zeropoint in time is set by this choice, our results are currently not
sensitive to it given the lateness of the epochs being surveyed in CSI, and so marginalize over $\gamma$ as a
nuisance parameter.

We grid each of the four unknowns, $\alpha$, $\beta$, $\gamma$, and $\sigma$, and compute four-dimensional posteriors.
As any {\it a priori\/} derivation of $\sigma$ is beyond the scope of this initial work, we also marginalized over it.


\begin{figure}
\centering
\includegraphics[width=0.37\textwidth]{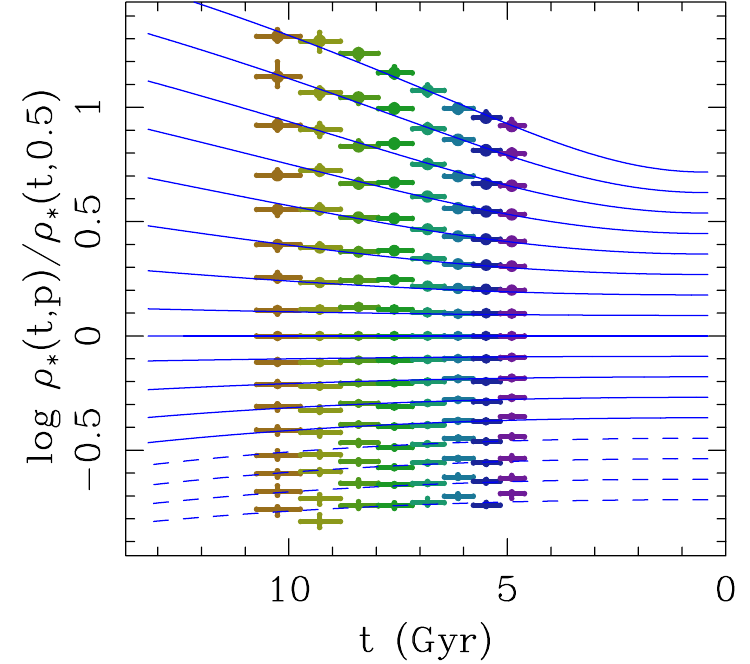}
\includegraphics[width=0.37\textwidth]{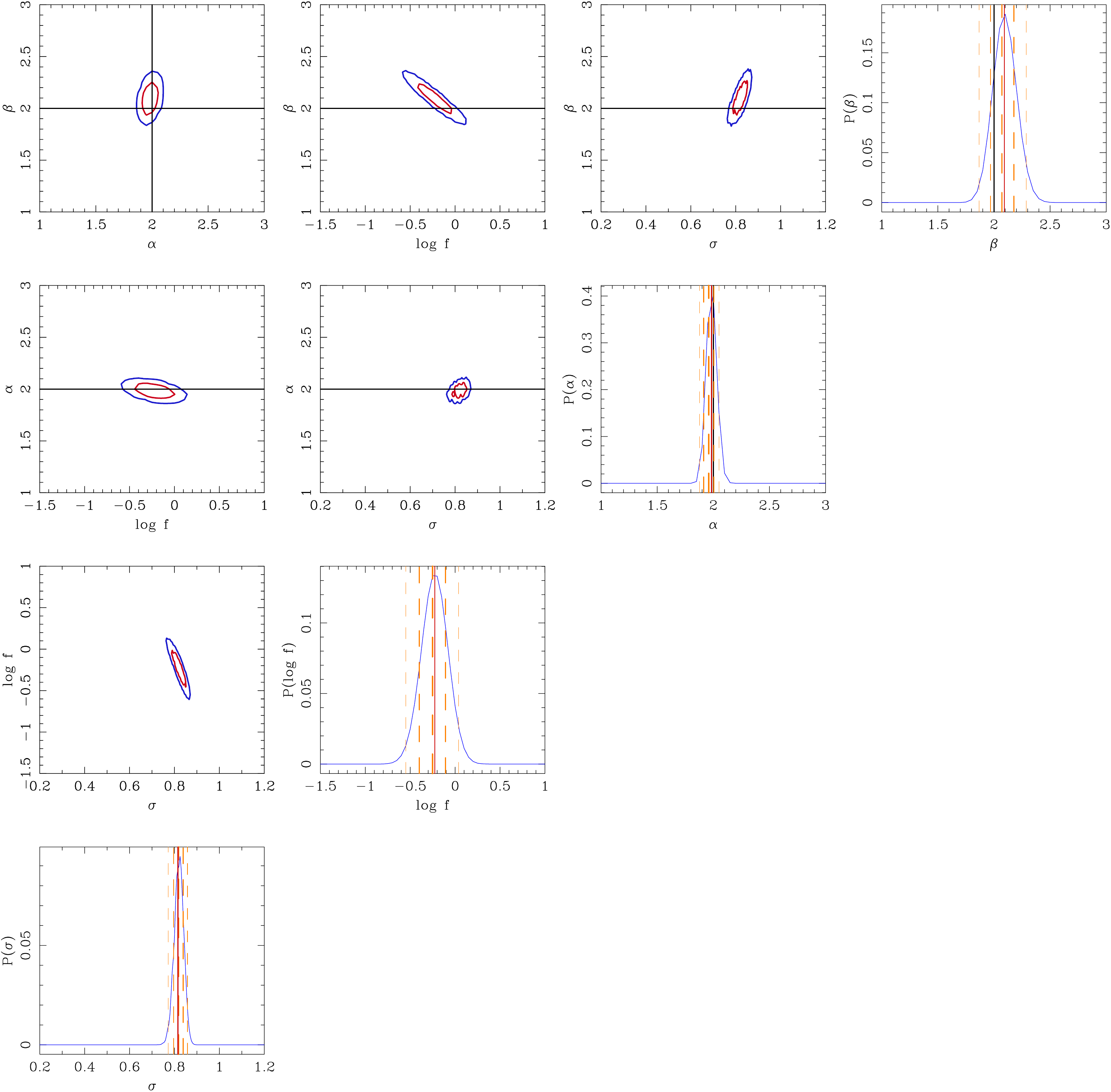}
\caption{(Top) A global fit of the power-law model for the mean growth of density percentiles in CSI is shown by the
blue solid lines. Dashed lines are shown at density percentiles at $\Phi^{-1}(p)< -1$, which were excluded from the
fitting to mitigate against potentially larger systematic uncertainties
in the measurements of low densities due to small number statistics. (Bottom) The contours for the 68\% and 95\%
confidence intervals in $\alpha$ and $\beta$ for the fit to the binned data shown at (Top).
\label{fig:Global}}
\end{figure}


Figure \ref{fig:Global}(top) shows one set of results from fitting the evolving densities at fixed
percentile, using the binned CSI data shown in Figure \ref{fig:DensityDist2}. Solid blue lines trace
the best-fit model for those percentiles used in the fitting. The lowest density percentiles, expected to be systematics-dominated, were excluded from the fit and their model curves are shown using the dashed blue lines. 
In Figure \ref{fig:Global}(bottom), 68\% and 95\% contours of the joint
posteriors on $\alpha$ and $\beta$ are shown, with marginalized confidence intervals of
$\alpha=1.96\pm 0.04$ and $\beta=2.07\pm 0.11$ ($68\%$). These inferences for the density- and
time-dependence of the mean growth of structure in the nonlinear regime do have some mild
sensitivity to the details of the redshift binning and to the samples of galaxies used in the
analysis.

\begin{figure}
\centering
\includegraphics[width=0.32\textwidth]{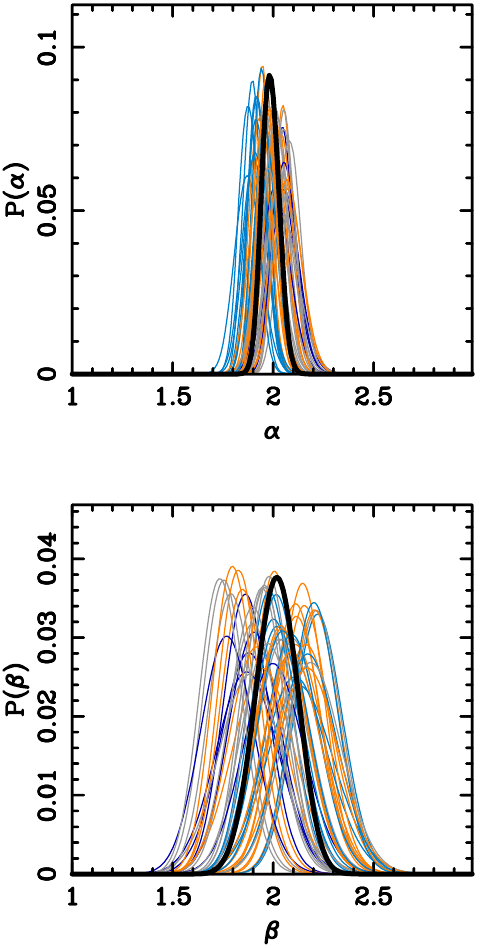}
\caption{(Top and Bottom) The marginalized posteriors for $\alpha$ and $\beta$ in thin coloured lines
from forty variants of slicing the CSI dataset, in stellar mass and redshift binning. Combining the
posteriors from the different slicings of the data, we have derived a combined set of posteriors for
$\alpha$ and $\beta$ shown by the thick black lines. We find that $\alpha=1.98 \pm 0.04$ and
$\beta=2.01\pm 0.11$, consistent with expectations for unabated gravitational collapse of early density fluctuations.
\label{fig:Marginalize}}
\end{figure}

We have varied the depth of the selection, the widths of the redshift slices, as well as the maximum redshift
of our study. We show marginalized posteriors for $\alpha$ and $\beta$ from $\sim 50$ such variants of
our analysis of the CSI data in Figure \ref{fig:Marginalize}. Most of these are not fully independent of each
other, but the combined posteriors, shown with thick black lines, do yield a small reduction in systematics related to the
redshift binning and stellar mass depth.

Using the one-dimensional posteriors in Figure \ref{fig:Marginalize}, we derive $\alpha=1.98\pm 0.04$ and $\beta=2.01\pm
0.11$. Recall from \S \ref{sec:frame}---and Equation \ref{eq:traj} in particular---that under the hypothesis that
density fluctuations grow through (Newtonian) gravitational collapse, the average growth rates of ensembles of early
density fluctuations should scale with the square of their initial density contrasts---an expectation now confirmed.
Secondly, mean growth trajectories were predicted to be quadratic in time (again, the time since the epoch of
decoupling). While the measurement uncertainties are larger for $\beta$ than for $\alpha$, the data also confirm that the
mean growth trajectories follow the quadratic time dependence expected for gravitational
collapse.\footnote{This mean dependence on time has been seen at the galaxy level as well---and leads
naturally to the observed ''main sequence of star formation'' and high-redshift stellar mass functions.
\citet{kelson2014} and \citet{kelson2016} explored these ramifications from the standpoint of galaxy growth as a
stochastic process.}
Together, these are the first direct comparisons of observational data on these scales to fundamental predictions for
nonlinear structure formation.

In the next section we discuss a few implications and side effects of our results.


\section{Implications}
\label{sec:imps}

For several decades, the most popular approaches to modeling and understanding the growth of structure have been
through Monte Carlo techniques like $N$-body simulations or through the use of halo merger trees. And while such
approaches offer the promise of fully self-consistent models of galaxy growth over a broad range of
astronomically interesting lengthscales, both internal and external to galaxies,
their richness and detail are not typically well-matched to the level of detail found in astronomical surveys.
This mismatch can limit the thoroughness and ease with which one can confront data with theory, but---more
importantly---it presents a substantial barrier to learning what astrophysics the data do speak to.
Exactly what information is present in observations of cosmological ensembles of galaxies? The answer to this
question remains elusive, as
fitting semi-analytical models to observations remains an underdetermined problem \citep[owing, e.g., to the
extensive covariances between parameters;][]{benson2014},

It is within such a context that one can view recent semi-empirical approaches to galaxy formation---such as by
\citet{peng2010} and \citet{peng2012}; or \citet{behroozi2013} and \citet{moster2013};
or \citet{gladders2013}, \citet{oemler2013} and \citet{abramson2016}; or by
\citet{kelson2014} and \citet{kelson2016}---as pieces of a nascent movement to identify the essential
content of observations, chiefly to uncover why some galaxies form stars more slowly or rapidly than others and at what
times.


The reduction in the numbers of assumptions or parameters in such galaxy evolution frameworks ought to have helped
observers glean no more than what their data could provide, but because such semi-empirical frameworks are
paradigmatically distinct \citep[see][]{abramson2016}, they credit different---and often incompatible---physics
as central to explanations for trends in the data. These incompatibilities are axiomatic, such that
the concerns of one framework may even be unintelligible---literally without any associated or interpretable meaning---to
the others.

And depending on whether one begins from the data---such as
starting with the {\it a priori\/} assumption of a narrow, empirically prescribed ``main sequence of star formation'' on
which all growing galaxies form stars as in \citet{peng2010} until they don't; or that there exist unique and monotonic
mappings between theoretical dark matter halo accretion histories and stellar mass growth \citep{moster2013,behroozi2013};
or the {\it a priori\/} adoption of the  Lilly-Madau plot
\citep{madau1996,lilly1996} as indicative of Hubble-timescale life cycles of galaxies as in \citet{gladders2013} or
\citet{abramson2016}; or whether one begins with {\it a priori\/} assumptions that the physical process of stellar mass
growth is simply a stochastic process \citep{kelson2014,kelson2016}---each conceptual framework yields strong qualitative
{\it and quantitative\/} statements about the life cycles of galaxies while reproducing many other aspects of galaxy-level data.

Here we attempt to explicitly bridge the gap between $N$-body simulations and those semi-empirical
conceptual---analytical---frameworks. By recognizing that astronomers are in the business of measuring differences
between {\it different\/} cosmologically representative volumes observed at different epochs, we
hold---operationally---that our data sets most accurately reflect volume averages of the complex, nonlinear dynamics
that undergird the
evolving density field (in the real universe as well as in cosmological simulations). Upon volume averaging the fluid
equations, we obtained relatively simple analytical expressions for the expected growth of early density
fluctuations---expressions confirmed by measurements of the distributions of local densities in CSI over 7 Gyr of
history.

With exact analytical prescriptions for the average growth of those density fluctuations transitioning to nonlinearity
at the start of star formation in hand, we can next ask how baryons follow or break from these expectations, and how
galaxy number counts follow or break from these expectations, to begin the work of asking why (and in which kinds of
density fluctuations) fewer and fewer baryons are turned into stars after $z\sim 2$, i.e. why and where the Lilly-Madau
plot turns over.

This new analytical approach to the growth of structure appears to have promise as a new conceptual framework for
modeling the galaxy in a statistical mechanical sense. That the MDPL2 simulations \citep{klypin2016,knebe2018} show
similar trends implies, too, that the catalogs of objects being produced by $N$-body simulations reflect the physical
processes encoded by the fluid equations and (Newtonian) gravity (as they should).
So while simulations ought to continue providing
detailed ramifications of physics on small---galaxy-level---scales, this new work suggests that a volume-averaged
analytical approach---encapsulating the general climate in which galaxies grow---should be seen as
mathematically complementary to Monte Carlo methodologies, while being potentially better matched operationally to
astronomical survey data.

However, if these results are seen stricly as a positive endorsement for standard approaches to the growth of structure, a
set of stark warnings lie beneath the surface. Because another lens with which to view these results is through the language
of the growth of structure as a horrible diffusion problem, though one where the particles do {\it not\/} experience stochastic changes
to their (growth) trajectories like those in an ideal gas. If growth trajectories were like particle trajectories in an
ideal gas, changes to accretion rates would be truly random, and mean growth rates, $\langle D\delta /Dt\rangle$ would
scale with $t^{1/2}$. Under such a scenario growth of density fluctuations would be ``Brownian,'' or ``Markovian,''
another way of saying that changes to accretion rates at a given time are always, and everywhere, independent of
previous ones.

But during nonlinear gravitational collapse, galaxies and halos---all matter density fluctuations---interact
with each other, imposing what may be called ``long-range dependence'' on each others' growth trajectories by some, or $1/f$ noise by many others.
The gravitational interaction between density fluctuations then occurs on all timescales, meaning that changes to
accretion rates become correlated with previous changes, {\it always and everywhere\/}. The more (and longer) the particles
interact with each other, the more the mean accretion rate, $\langle D\delta /Dt\rangle$, grows faster than $t^{1/2}$
through positive reinforcement.  The result is that a process of matter accretion scales as $t^H$ where $H$ is known
as the ``Hurst parameter.''

In \S \ref{sec:frame} we derived $\langle D\delta/Dt\rangle\propto t$, implying that the growth of structure acts like a
nonnegative stochastic process with $H=1$ (or a process with a $1/f$ noise spectrum identically). Intriguingly, this is the same kind of process that was identified at
the galaxy level by studying the main sequence of star formation and the stellar mass function
\citep[e.g.][]{kelson2016}, which both behave as if they were governed by a process with $H=1$.

An important implication immediately follows; quoting \cite{mandelbrot1968}, about the properties of a stochastic process $X(t,\ldots)$:
\begin{lquote}
In analyzing time series $X(t,\ldots)$... it is customary to search for a
decomposition into a ``linear trend component'' and an ``oscillatory component.'' The former usually ... is
interpreted as due to major ``causal'' changes in the mechanism generating $X(t,\ldots)$. The latter, on the
contrary, is taken to be an ``uncontrollable'' stationary process, hopefully free of low-frequency components.
\end{lquote}
\begin{lquote}
It is obvious that, in the case of [$\ldots$] $H\ne 1/2$, difficult statistical problems are raised by
the task of distinguishing the linear trend $\Delta t$ from the nonlinear ``trends'' just described.
\end{lquote}
That is, when $H>1/2$, the noise itself---and long-term correlations in the noise---imposes long-term
trends through (positive) reinforcement. In the case of nonlinear structure formation, gravity---and gravity alone---provides
that reinforcement. \cite{mandelbrot1968} end the above passage by stating that such stochastic processes fall
{\it ``outside the usual dichotomy between causal trends and random perturbations.''\/} Stated even more directly:
standard tools for interpreting astronomical observations of galaxy ensembles are not appropriate for extracting
historical information about individual galaxies.

As observers, astronomers are used to looking for signal by smoothing over what appears to be noise. Galaxy evolution
as a practised field of study has largely been an effort to trace medians (or means) of galaxy properties with
redshift/time, to smooth over the deviations from those medians as if those deviations were the ``uncontrollable noise.''
Decades have been spent looking further into the distant past in order to continue measuring that ``linear trend
component''---the mean evolution of scaling relations---under the assumption that these measurements constrain
fundamental laws of astrophysics in some galaxy formation theory that is both accessible and deterministic. The tools to
exploit astronomical data by other---and more meaningful---means are practically non-existent, due to the fact that
the ``linear trend component'' and ``the noise'' both arise from the same underlying process, and in equal measure.
As a result, the development of new tools to model distributions should be seen as paramount if progress is to be made in
extracting meaning from cross-sectional studies of galaxies---not from the evolution of medians of the data, but from the ``uncontrollable noise''
that itself is the signal.

Until now, however, no path to generating accurate models of the nonlinear growth of structure---outside of $N$-body
simulations---had seemed possible. But by shifting from an Eulerian coordinate system for the growth of structure, to a
Lagrangian one, we have begun assembling an analytical framework for galaxy formation that more closely
aligns with the cross-sectional nature of astronomical studies---for the evolving density field, as well as for the distinct objects that are
observed at late times ($z\simlt 10$). By doing so,
we aim to explicitly connect the early power spectrum to the evolving content of baryons and stellar mass, and in
a manner connected to the growth of the objects we directly observe and study.

So it is with a sense of optimism that we look forward, to pushing this new conceptual approach to its logical end.
Derivations of analytical expectations, such as those in \S \ref{sec:frame} and elsewhere \citep{kelson2014,kelson2016},
should be taken as starting points for new {\it analytical\/} models of the macroeconomy of galaxy formation. For
example, the volumetric averaging of stellar mass growth can be coupled with the diminishing supply of cool baryons,
and, e.g., an ever increasing presence of entropy \citep[see, e.g.][]{kelson2016}, to construct ordinary differential
equations for the evolution of the star formation rate density in different ensembles of early density
fluctuations---different $\delta_p$---with predictions for galaxy ensembles, their stellar populations, or even their
heavy element yields, as functions of late-time density.

Surely when this picture is complete, it will not only be a robust description of the evolving density field that feeds
galaxy growth, but should provide new insights into its macroscopic galaxy content as well---with an exactitude
thought only approachable through Monte Carlo techniques. More importantly though, by bypassing much of the fine-tuning
of deterministic parameters that can complicate other theoretical approaches, we will have confidence that whatever is
implied by such models will be an abstraction of exactly what our data have been trying to say all along.


\section*{Acknowledgements}

The team acknowledges the generous support of the technical, administrative, engineering, custodial, and scientific
staff of the Carnegie Institution---in California, in Chile, and in Washington, D.C.---in making large, innovative
projects like CSI possible.  Without well-supported environments that can foster slow, deliberate, and cautious
interpretations of carefully curated data sets, it would not be possible to take a fresh look at the underpinnings of
the galaxy formation ecosystem with one eye on theory, another on astroinformatics, and yet another on galaxies
themselves. The authors also acknowledge Dr. J. Crane's efforts in gluing the Uniform Dispersion Prism and Dr. S.
Burles's efforts in creating and assembling the original Low Dispersion Prism. The first author thanks his collaborators
for indulging in digressions such as these, which are finally beginning to coalesce around the information content of
our survey, not to mention the rest of the extragalactic sky. C. Gauss did some cool stuff.
The anonymous referee is also acknowledged for prodding us to rethink connecting our approach to previous work.


\bigskip
\bsp	
\label{lastpage}
\end{document}